\documentclass[fleqn,usenatbib]{mnras}

\usepackage{newtxtext,newtxmath}

\usepackage[T1]{fontenc}

\usepackage{graphicx}	
\usepackage{amsmath}	
\usepackage{listings}
\usepackage{xcolor}
\usepackage{siunitx}
\usepackage{booktabs}

\definecolor{codegreen}{rgb}{0,0.6,0}
\definecolor{codegray}{rgb}{0.5,0.5,0.5}
\definecolor{codepurple}{rgb}{0.58,0,0.82}
\definecolor{backcolour}{rgb}{0.95,0.95,0.92}

\lstdefinestyle{mystyle}{
    backgroundcolor=\color{backcolour},   
    commentstyle=\color{codegreen},
    keywordstyle=\color{magenta},
    numberstyle=\tiny\color{codegray},
    stringstyle=\color{codepurple},
    basicstyle=\ttfamily\footnotesize,
    breakatwhitespace=false,         
    breaklines=true,                 
    captionpos=b,                    
    keepspaces=true,                 
    numbers=left,                    
    numbersep=5pt,                  
    showspaces=false,                
    showstringspaces=false,
    showtabs=false,                  
    tabsize=2
}

\newcommand{\heyoka}{\texttt{heyoka}}

\title[High-order Taylor methods]{Revisiting high-order Taylor methods for astrodynamics and celestial mechanics}

\author[F. Biscani et al.]{
Francesco Biscani,$^{1}$\thanks{E-mail: bluescarni@gmail.com}
Dario Izzo$^{2}$
\\
$^{1}$Center for Astronomy (ZAH), University of Heidelberg / Max Planck Institute for Astronomy, Heidelberg, Germany\\
$^{2}$Advanced Concepts Team, European Space Agency (ESA), Noordwijk, The Netherlands
}

\date{Accepted XXX. Received YYY; in original form ZZZ}

\pubyear{2021}

\begin{document}
\label{firstpage}
\pagerange{\pageref{firstpage}--\pageref{lastpage}}
\maketitle

\begin{abstract}
We present \heyoka{}, a new, modern and general-purpose implementation of Taylor's integration method for the numerical solution of
ordinary differential equations. Detailed numerical tests focused on difficult high-precision gravitational problems in astrodynamics and celestial
mechanics show how our general-purpose integrator is competitive with and often superior to state-of-the-art specialised
symplectic and non-symplectic integrators in both speed and accuracy. In particular, we show how Taylor methods are capable of
satisfying Brouwer's law for the conservation of energy in long-term integrations of planetary systems over billions
of dynamical timescales. We also show how close encounters are modelled accurately during simulations of the formation of the Kirkwood
gaps and of Apophis' 2029 close encounter with the Earth (where \heyoka{} surpasses the speed and accuracy of
domain-specific methods).
\heyoka{} can be used from both C++ and Python, and it is publicly available as an open-source project.
\end{abstract}




\section{Introduction}
The search for efficient and accurate numerical methods for the solution
of ordinary differential equations (ODEs) in celestial mechanics
and dynamical astronomy has a long history.
Because gravitational systems of scientific interest exhibit a large variety of properties
and behaviours, a multitude of integration methods
targeting specific domains has been developed over the years.

In the context of Solar System dynamics, for instance, long-term integrations
are often performed using symplectic integrators,
which have the desirable property of guaranteeing the conservation of (a slightly perturbed version of) the
energy of the system \citep{de1956methods,ruth1983canonical}.
In stellar dynamics, Hermite integration schemes with hierarchical
time steps are particularly popular \citep{aarseth2003gravitational}. For
large-scale gravitational system (e.g., in galactic and cosmological simulations),
in order to tame
the computational complexity,
fast methods for the computation of the gravitational force
are often paired to low-order integrators \citep{barnes1986hierarchical, greengard1987fast}.

Although Newtonian gravity is usually the dominant force in astrodynamics and celestial mechanics, it
is often crucial for numerical integrators to be able to account for non-Newtonian and non-conservative interactions as well.
The Yarkovsky effect, for instance, is understood to be one of the main drivers pushing asteroids
from the belt onto near-Earth trajectories \citep{gladman1997dynamical}. The interaction with the proto-planetary
disk, which can be modelled as a drag, plays a crucial role in the evolution of planets and planetesimals \citep{capobianco2011planetesimal}.
Relativistic effects, which are important in high-accuracy studies of the future evolution of the Solar
System, are usually modelled as velocity-dependent forces in the framework of post-Newtonian approximations
\citep{laskar2011la2010}.
A further important aspect to consider carefully is the ability of an integration method to accurately resolve
the dynamics over a wide range of timescales.
In the context of planet population synthesis, for instance, the long-term orbital evolution
of a system of planetary embryos is punctuated by close encounters and collisions which
gradually lead to coalescence into planets  \citep{mordasini2009extrasolar}.
Accurately modelling these close encounters necessitates the use of adaptive timestepping
and/or switching to a dedicated high-precision integration
method for the duration of the encounter \citep{1997DPS....29.2706C}.

In this paper we revisit Taylor's method, one of the oldest numerical procedures for the solution of ODEs, with
a focus on high-precision astrodynamical applications. Despite its appealing features,
Taylor's method is not widely adopted in the astrodynamical community, due, we believe,
to its implementation difficulty and the suboptimal ergonomics of existing packages (which
usually require the use of preprocessors or translators, see \citet{chang1994atomft, jorba2005, abad2012algorithm}).
Our new implementation of Taylor's method, called \heyoka{}, is general-purpose, it is usable directly from the C++ and Python languages
and it does not require any intermediate translation step, relying instead on a just-in-time
compilation approach. Innovative features include support for
extended precision arithmetics and the ability to exploit the vector instructions of modern
processors to provide a substantial throughput increase with respect to scalar computations.

From the scientific point of view, we will show how our general-purpose implementation of
Taylor's method outperforms state-of-the-art domain-specific
symplectic and non-symplectic integrators in a variety of challenging problems
in gravitational dynamics. In particular, we will show for the first time how Taylor's
method is capable of respecting Brouwer's law for the conservation of energy in high-accuracy long-term
integrations of planetary systems over billions of dynamical timescales. We will also show
how, at the same time, the general-purpose adaptive timestepping strategy adopted in our implementation
is capable of accurately modelling close encounters and high-eccentricity orbits.

In many ways, our Taylor integrator is close in spirit to the \texttt{IAS15} integrator
presented in \citet{rein2015ias15}, to which our implementation will be frequently compared
throughout the paper. Like \texttt{IAS15}, our integrator is a non-symplectic adaptive-timestep scheme which is however
capable of preserving energy as well as symplectic integrators in high-accuracy setups. Moreover,
like \texttt{IAS15}, our integrator is also capable of accurately modelling non-conservative
forces and close gravitational flybys, two areas in which fixed-timestep symplectic integrators
encounter difficulties. Unlike \texttt{IAS15}, \heyoka{} also provides a dense output without any added complexity.
Furthermore, it is a variable order method, allowing a seamless extension of its properties to extended precision arithmetics.  

The paper is organised as follows:
in Section \ref{sec:taylor_overview}, we give a brief overview of the salient features of Taylor methods;
in Section \ref{sec:impl}, we describe the details of our implementation; in Section \ref{sec:tests}, we test our Taylor integrator in a variety of challenging gravitational problems;
in Section \ref{sec:limitations}, we highlight a few limitations and shortcomings of Taylor methods, and show how a Taylor integrator could benefit from being specialised for specific purposes.

\heyoka{} is freely available as an open-source project at \url{https://github.com/bluescarni/heyoka}.

\section{Overview of Taylor methods}
\label{sec:taylor_overview}
In this section, we will give a brief overview of Taylor methods and their most important properties.
For a more detailed discussion which includes a historical perspective, we refer to \citet{jorba2005}.

We consider systems of ordinary differential equations (ODE) in the explicit form
\begin{equation}
\boldsymbol{x}'\left( t \right)=\boldsymbol{F}\left(t, \boldsymbol{x}\left( t \right) \right),\label{eq:ode_00}
\end{equation}
where $t$ is the independent variable and the prime symbol denotes differentiation with
respect to $t$. Given a set of initial conditions for $t=t_0$,
\begin{equation}
\boldsymbol{x}_0=\boldsymbol{x}\left(t_0\right),\label{eq:ode_ic_00}
\end{equation}
Taylor methods approximate the solution of \eqref{eq:ode_00} for $t=t_1$ as the
truncated Taylor expansion of the solution around $t=t_0$,
\begin{equation}
\boldsymbol{x}\left( t_1 \right) = \boldsymbol{x}_0 + \boldsymbol{x}'\left(t_0\right)h
+\frac{1}{2}\boldsymbol{x}''\left(t_0\right)h^2+\ldots+\frac{\boldsymbol{x}^{\left( p \right)}\left(t_0\right)}{p!}h^p,\label{eq:trunc_00}
\end{equation}
where $h = t_1 - t_0$ is the integration timestep, $\boldsymbol{x}^{\left( n \right)}\left( t \right)$ denotes the derivative of order $n$
of $\boldsymbol{x}\left( t \right)$, and $p$ is the \emph{order} of the Taylor method. After defining
\begin{equation}
a^{\left[ n \right]}\left( t \right) = \frac{1}{n!}a^{\left( n \right)}\left( t \right)\label{eq:n_diff_00}
\end{equation}
as the \emph{normalised derivative} of order $n$ of $a\left( t \right)$, eq. \eqref{eq:trunc_00} can be rewritten in a more
compact form as
\begin{equation}
\boldsymbol{x}\left( t_1 \right) = \sum_{n=0}^p \boldsymbol{x}^{\left[ n \right]} \left(t_0\right) h^n.\label{eq:taylor_poly_00}
\end{equation}

Clearly, Taylor methods belong to the family of explicit integration methods since the update rule detailed in
eq. \eqref{eq:taylor_poly_00} is an explicit equation for $\boldsymbol{x}\left( t_1 \right)$.
As such, Taylor methods will have issues with problems having a certain degree of stiffness.
\citet{abad2012algorithm} show how Taylor methods are A-stable in the limit $p \to \infty$ (see Proposition 2.1),
which in practice means that high-order Taylor methods can be successfully used to integrate
moderately-stiff problems. It is also possible to formulate implicit versions of the Taylor integration
scheme \citep{kirlinger1991implicit}. The resulting method, however, necessitates some predictor
scheme which ultimately will result in a slower procedure, albeit able to solve also highly-stiff problems.

Note also how Taylor methods directly provide, via eq. \eqref{eq:taylor_poly_00}, a high-precision approximation
of the solution of the ODE system within a timestep. In other words, Taylor integrators provide \emph{dense output}
at no additional cost. Moreover, the dense output provided by a Taylor integrator is not the result of an
interpolation (as commonly done in other integration methods), but rather it is a high-fidelity approximation
of the solution via the direct computation of its Taylor series.

The efficient computation of the normalised derivatives
$\boldsymbol{x}^{\left[ n \right]} \left(t_0\right)$ in eq. \eqref{eq:taylor_poly_00}
is one of the main problems to solve in the implementation of a Taylor integrator.
In principle, using the first-order derivatives
from \eqref{eq:ode_00} it is possible to derive explicit expressions for $\boldsymbol{x}^{\left[ n \right]} \left(t_0\right)$
as functions of $\boldsymbol{x}_0$ via elementary calculus. This approach, however, suffers from serious shortcomings:
\begin{itemize}
    \item the formul\ae{} of the explicit expressions depend on the ODE system begin integrated (i.e., they cannot be computed
    once and for all);
    \item obtaining the explicit expressions requires the calculation of a large number of high-order derivatives, which, if done by hand, is
    a tedious and error-prone process;
    \item in general, the complexity of explicit expressions for the derivatives grows exponentially with the order $n$.
\end{itemize}
While the first two issues can be ameliorated at the implementation level (e.g., through the use of a symbolic manipulation
library), the complexity issue limits the usefulness of Taylor methods at high orders. Luckily, this obstacle can be overcome
via a process of \emph{automatic differentiation} as detailed in the next section.

\subsection{Automatic differentiation}
\label{subsec:AD}
In the context of Taylor integration, automatic differentiation (AD) is a method to compute iteratively the values of high-order derivatives
starting from the values of the derivatives at orders 0 and 1. The fundamental idea behind this is that nontrivial mathematical expressions can be
seen as compositions of simple expressions whose $n$-th order derivatives have explicit formul\ae{}. In order to illustrate
the process, let us consider the simple ODE system representing the Van der Pol oscillator:
\begin{equation}
\begin{cases}
x' = y\\
y' = \left( 1 - x^2\right)y-x
\end{cases}
\label{eq:vdp_00}
\end{equation}
By introducing the following ancillary definitions,
\begin{equation}
\begin{aligned}
    u_1 &= x,\\
    u_2 &= y,\\
    u_3 &= u_1u_1,\\
    u_4 &= 1-u_3,\\
    u_5 &= u_4u_2,\\
    u_6 &= u_5-u_1,
\end{aligned}
\label{eq:u_defs_00}
\end{equation}
the ODE system \eqref{eq:vdp_00} can be rewritten as
\begin{equation}
\begin{cases}
x' = u_2\\
y' = u_6
\end{cases}
\label{eq:vdp_01}
\end{equation}
Because all the definitions in \eqref{eq:u_defs_00} consist of either identities or simple arithmetic operations, it is possible to write
explicit formul\ae{} for the normalised derivatives of the $u_i\left( t \right)$ variables. Specifically:
\begin{enumerate}
    \item if $c\left( t \right) = a\left( t \right)\pm b\left( t \right)$, then:
    $c^{\left[ n \right]}\left( t \right) = a^{\left[ n \right]}\left( t \right) \pm b^{\left[ n \right]}\left( t \right)$;
    \item if $c\left( t \right) = a\left( t \right)  b\left( t \right)$, then:
    $c^{\left[ n \right]}\left( t \right) = \sum_{j=0}^n a^{\left[ n - j \right]}\left( t \right) b^{\left[ j \right]}\left( t \right)$ (Leibniz's rule);
    \item if $c\left( t \right) = \frac{a\left( t \right)}{b\left( t \right)}$, then: \newline
    $c^{\left[ n \right]}\left( t \right) = \frac{1}{b^{\left[ 0 \right]}\left( t \right)}\left[a^{\left[ n \right]}\left( t \right) - \sum_{j=1}^n b^{\left[ j \right]}\left( t \right)c^{\left[ n - j \right]}\left( t \right) \right]$.
\end{enumerate}
See \citet{jorba2005, barrio2005performance, haro2008automatic} for derivations and additional AD rules for various
elementary functions (e.g., exponentiation, trigonometric functions, logarithm, exponential, etc.).

For instance, if we need to compute the second-order normalised derivative of $y$,
using \eqref{eq:vdp_01} and \eqref{eq:n_diff_00} we can see that
\begin{equation}
y^{\left[ 2 \right]} = \frac{1}{2} u^{\left[ 1 \right]}_6.
\end{equation}
From the definition of $u_6$ in \eqref{eq:u_defs_00}, we can write
\begin{equation}
u^{\left[ 1 \right]}_6 = u^{\left[ 1 \right]}_5 - u^{\left[ 1 \right]}_1.
\end{equation}
On the other hand, from the definition of $u_5$, via Leibniz's rule we can write
\begin{equation}
u^{\left[ 1 \right]}_5 = u^{\left[ 1 \right]}_4u^{\left[ 0 \right]}_2 + u^{\left[ 0 \right]}_4u^{\left[ 1 \right]}_2,
\end{equation}
and similarly we can recursively write the derivatives of $u_4$ in terms of the derivatives of $u_3$, and so on.

In other words, using AD it is possible to compute derivatives of order $n$ of the state variables
$x$ and $y$ via the composition of lower-order derivatives of the $u_i$ variables. Crucially, because the AD rules have a complexity which
is at most linear with the order of differentiation $n$, the computation of the $n$-th order derivative of a state variable
has at most quadratic complexity. This is of course a substantial improvement over the
exponential complexity of the explicit expressions for the derivatives.

\subsection{Choice of order and timestep}
\label{subsec:timestep}
Contrary to fixed-order methods, in a Taylor integrator one can choose
both the order $p$ \emph{and} the integration timestep $h$. Because different combinations of $p$ and $h$
can yield the same final integration error, there is a considerable flexibility in the way $p$ and $h$
can be chosen. In our work we follow the prescriptions from \citet{jorba2005}, which we will summarise here briefly.

\citet{jorba2005}, building on previous work by \citet{simo2001global},
show that, in the context of Taylor methods, if the objective is the minimisation
of the number of floating-point operations per unit of integration time (i.e., the maximisation
of the performance of the integrator), then there exists an optimal choice for the order $p$.
Specifically, \citet{jorba2005} provide the following formula for the optimal Taylor order $p$:
\begin{equation}
p = \left\lceil -\frac{1}{2}\log \epsilon_m + 1 \right\rceil,\label{eq:opt_order_00}
\end{equation}
where $\left\lceil \right\rceil$ is the ceiling function and $\epsilon_m$ is the maximum absolute error
that we want to allow when truncating a Taylor series. This result stems from the fact
that the cost of computing the coefficients of the Taylor series via AD is quadratic
in the Taylor order (see \S3.1 and \S3.2 in \citet{jorba2005} for details). Thus,
even though, e.g., Taylor orders higher than $p$ would allow longer integration timesteps
while achieving the same absolute error $\epsilon_m$, the quadratic increase in
complexity for the computation of the additional Taylor coefficients would ultimately lead
to a slower integrator.

\citet{jorba2005} then distinguish two ways in which the user can specify
the desired integration accuracy:
an absolute tolerance $\epsilon_a$ and a relative tolerance $\epsilon_r$. The value for $\epsilon_m$ in eq. \eqref{eq:opt_order_00} is then set to $\epsilon_a$
when $\epsilon_r \left\lVert \boldsymbol{x}_0 \right\rVert_\infty \leq \epsilon_a$, and to $\epsilon_r$ otherwise. We denote with
$ \left\lVert \cdot \right\rVert_\infty$ the infinity norm of a vector (i.e., the maximum magnitude of its components).
In our implementation, however, we impose $\epsilon_r = \epsilon_a$ (that is, the user specifies only one tolerance value).
Thus, a unique optimal Taylor order $p$ derived from  eq. \eqref{eq:opt_order_00} is used throughout the integration. We do, however, switch between an absolute and relative error control mode in correspondence of $\left\lVert \boldsymbol{x}_0 \right\rVert_\infty = 1$.

In absolute error control mode (i.e., $\left\lVert \boldsymbol{x}_0 \right\rVert_\infty \leq 1$ in our implementation), \citet{jorba2005}
introduce the quantities
\begin{equation}
\rho_m^{\left( j \right)} = \left( \frac{1}{\left\lVert \boldsymbol{x}^{\left[ j \right]} \left(t_0\right) \right\rVert_\infty}  \right)^\frac{1}{j}
\label{eq:rho_m_abs_00}
\end{equation}
and
\begin{equation}
\rho_m = \min \left\{ \rho_m^{\left( p-1 \right)}, \rho_m^{\left( p \right)} \right\}.\label{eq:rho_min}
\end{equation}
$\rho_m$ is essentially an estimation of the smallest radius of convergence among the Taylor
series of all the state variables consequence of the Cauchy-Hadamard theorem. The integration timestep
is then computed as
\begin{equation}
h = \frac{\rho_m}{e^2}\exp \left( -\frac{0.7}{p-1} \right),\label{eq:ts_00}
\end{equation}
where $\exp \left( -\frac{0.7}{p-1} \right)$ is a safety factor depending on the order $p$ (see \S3.2.1 in \citet{jorba2005}).

In relative error control mode (i.e., $\left\lVert \boldsymbol{x}_0 \right\rVert_\infty > 1$ in our implementation),
the $\rho_m^{\left( j \right)}$ quantities are defined, instead of as in \eqref{eq:rho_m_abs_00}, as
\begin{equation}
\rho_m^{\left( j \right)} = \left( \frac{\left\lVert \boldsymbol{x}_0 \right\rVert_\infty}{\left\lVert \boldsymbol{x}^{\left[ j \right]} \left(t_0\right) \right\rVert_\infty}  \right)^\frac{1}{j},
\label{eq:rho_mj_rel}
\end{equation}
so that $\rho_m$ becomes an estimation of the smallest radius of convergence rescaled by the largest
value in the state vector. The timestep $h$ is then computed, like in absolute error control mode, via
eqs. \eqref{eq:rho_min} and \eqref{eq:ts_00}.
Note that the dependency of $\rho_m^{\left( j \right)}$ on $\left\lVert \boldsymbol{x}_0 \right\rVert_\infty$
in eq. \eqref{eq:rho_mj_rel} is weak, especially at high Taylor orders, due to the presence of the $1/j$ exponent.

It can be shown that, if $p$, $\rho_m$ and $h$ are chosen in the manner explained above, then
the error resulting from the truncation of the Taylor series \eqref{eq:taylor_poly_00} is bounded, either in an absolute or
relative sense, by the user-defined tolerance (see Proposition 3.4 in \citet{jorba2005}).

Although the error control approach of \citet{jorba2005} has proved to be effective in our tests
and numerical experiments, other choices are possible and might be more appropriate in certain circumstances.
The error control in \texttt{ATOMFT} \citep{chang1994atomft}, for instance, uses interval arithmetic
to provide mathematically certified error bounds. \texttt{TIDES} \citep{abad2012algorithm}, on the other hand,
implements both a strategy based on asymptotic estimates (like the one we described) and another strategy
which takes into account the history of the system.

In order to cater for different usage scenarios, our implementation provides both an easy-to-use
integrator interface based on the automatic error control described in this section, and lower-level
functions (such as the computation of the jet of derivatives and a state propagator) that
can be used to implement alternative timestepping strategies.

Note that the adaptive timestep scheme described in this section does not require
any warmup, it does not rely on any domain-specific knowledge and it depends solely on the values
of the derivatives at the beginning of the timestep. Thus, when using the timestep deduction scheme
described in this section, a Taylor integrator is a single-step method.

\subsection{Computations at high accuracy}
\label{sec:comp_high_acc}
One important feature of Taylor methods is that, contrary to fixed-order methods, it is not necessary
to decrease the timestep $h$ in order to reduce the integration error. Instead, one can simply increase the Taylor
order $p$: because the number of floating-points operations
necessary to perform an integration timestep scales quadratically with the Taylor order $p$
(as we mentioned in \S\ref{subsec:AD}), increasing $p$ is a more efficient way of reducing the error than decreasing the
timestep $h$. Let us illustrate this point with a concrete example.

If one wants to achieve machine precision when operating in IEEE double-precision arithmetic, then the
tolerance should be set to $\sim 2.2\times 10^{-16}$. According to \eqref{eq:opt_order_00}, this results
in a Taylor order $p=20$. When using IEEE quadruple-precision arithmetic, in order to achieve
machine precision, the tolerance should be set to $\sim 10^{-34}$, which yields a Taylor order of $p=40$. In other words,
switching from double to quadruple precision requires doubling the Taylor order, and, correspondingly,
the number of floating-point operations necessary to perform a single step increases by a factor of $2^2=4$. On the other
hand, when using an integrator with fixed order, say, 15,
then in order to achieve machine precision when moving from double to quadruple precision, one needs to
decrease the timestep $h$ (and thus increase the number of floating-point operations)
by a factor of $10^\frac{34-16}{15+1} \sim 13$.

\section{Implementation}
\label{sec:impl}
Whereas in the previous section we briefly discussed the theoretical underpinnings of Taylor methods,
in this section we will present the details of our implementation.

Existing software packages implementing Taylor methods usually consist of a translator program
which accepts as input a human-readable description of the ODE system to be integrated, and returns
as output a source code file (usually in Fortran or C) providing a timestepper function
built on top of the automatic-differentiation machinery presented in \S\ref{subsec:AD}.

In the \texttt{taylor} package \citep{jorba2005}, for instance, the user is expected to
write the ODEs into a text file following a custom grammar. The text file is then processed
by the \texttt{taylor} translator to produce a set of C files implementing the timestepper. \texttt{ATOMFT}
\citep{chang1994atomft} works in a similar fashion, but producing source code in Fortran
rather than in C. The \texttt{TIDES} package \citep{abad2012algorithm,abad2015automatic} eschews the use of a custom
grammar for the description of the ODE system in favour of the
Mathematica language \citep{wolfram1999mathematica} or the \texttt{SAGE} computer algebra system \citep{sagemath}.
Like \texttt{taylor} and \texttt{ATOMFT}, \texttt{TIDES} then translates the symbolic description of the ODE
system into a set of C or Fortran files to be compiled separately.

Our implementation, instead, removes the need of a translator program altogether in favour of an implementation
fully contained within a single C++ library. Instead of relying on an auxiliary language for the
representation of the ODE system, we employ a small, self-contained
symbolic expression system which allows to represent ODEs directly in the C++ grammar. The ODEs
are then decomposed into elementary expressions in the manner explained in \S\ref{subsec:AD}, resulting
in a sequence of AD formul\ae{} which are assembled and compiled just-in-time (JIT) via the LLVM library
\citep{lattner2004llvm} to synthesise a complete ODE integrator.

Although the core of our Taylor integrator is implemented in C++, we also provide a Python
interface which enables interactive access to the main functionality of our library.
\heyoka{}'s source code can be freely downloaded from \url{https://github.com/bluescarni/heyoka}.

\subsection{Representing symbolic expressions}
\begin{figure}
  \centering
  \includegraphics[width=.7\columnwidth]{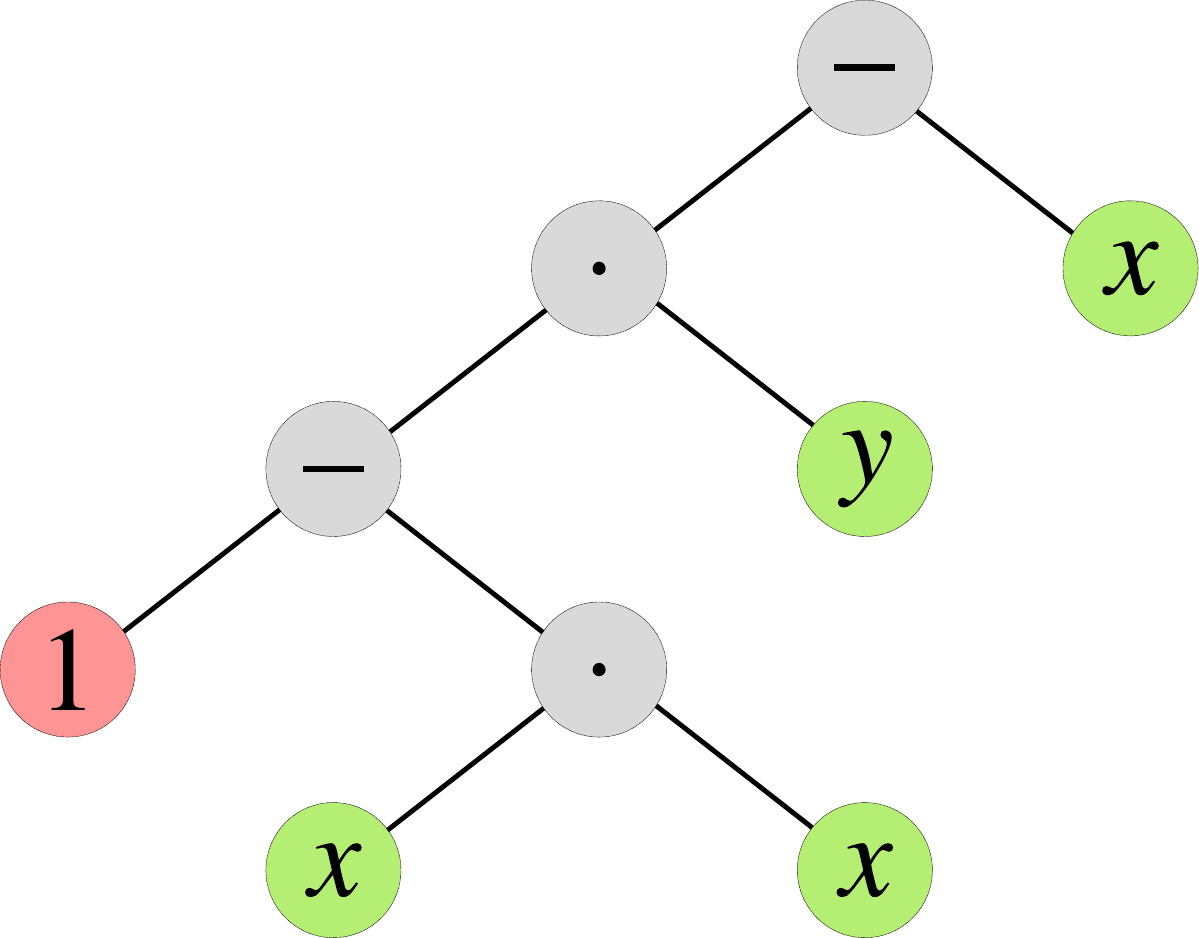}
  \caption{Abstract syntax tree (AST) of the mathematical expression $\left( 1 - x^2\right)y-x$, representing
  the right-hand side of the second equation of the Van der Pol ODE system (see eq. \eqref{eq:vdp_00}).
  The internal nodes (in this specific case, binary operators) are visualised in gray, while the leaf nodes are visualised in red (constants)
  and green (symbolic variables).}
  \label{fig:ast_00}
\end{figure}

In our implementation we represent symbolic expressions using a simple abstract syntax tree (AST)
in which the internal nodes can be binary operators or $n$-ary functions, and the leaf nodes
can be variables or constants. Figure \ref{fig:ast_00} displays a graphical representation
of the AST of the right-hand side of the second equation of the Van der Pol ODE system (see eq. \eqref{eq:vdp_00}).

Thanks to C++ features such as operator overloading and user-defined literals,
the construction of the AST of an expression is hidden
behind simple mathematical notation. For instance, the right-hand side of the second equation
of the Van der Pol system can be defined as follows:
\begin{lstlisting}[language=C++]
// Create the symbolic variables.
auto x = "x"_var;
auto y = "y"_var;

// Define an expression. The constant 1
// is created in double precision via the
// use of the "_dbl" literal.
auto ex = (1_dbl - x*x) * y - x;

// Print the expression to screen.
std::cout << ex << '\n';
\end{lstlisting}
In addition to the basic arithmetic operations, our
expression system supports several special functions.
More functions can be added via an extension API without modifying the
library's code.

The expressions' tree structure allows to carry out the
decomposition necessary to implement the automatic differentiation formul\ae{} in
a straightforward fashion via a simple depth-first traversal of the AST. For instance,
in Figure \ref{fig:ast_00} we can see that the depth-first traversal of the Van der Pol equation's AST
encounters $x \cdot x$ as the first elementary subexpression. $u_3 = x \cdot x$ is also the definition
of the $u_3$ variable in the decomposition \eqref{eq:u_defs_00}. The next subexpression encountered
in the depth-first traversal is $1-x\cdot x = 1-u_3$, which is the definition of the $u_4$ variable
in the decomposition \eqref{eq:u_defs_00}. Similarly, continuing the traversal produces the definitions of
$u_5$ and $u_6$ in terms of the preceding $u_i$ variables.

During the decomposition of an ODE system, it is fairly common for the same elementary subexpressions to appear multiple
times. This is the case, for instance, in N-body systems, where the distance between two particles, $r_{ij}$,
appears in the differential equations of both particle $i$ and $j$. In order to avoid redundant computations,
we detect and remove duplicate expressions in the decomposition via a process of common subexpression
elimination.

\subsection{Supported floating-point types}
\label{subsec:supp_fp_types}
As we mentioned earlier, Taylor integration methods are particularly well-suited 
when operating at high precision, because increasing the accuracy by raising the Taylor order $p$ is more
computationally efficient
than reducing the integration timestep $h$. In order to enable high-accuracy computations, our implementation
can use different floating-point types when synthesising an ODE integrator. Specifically, the following
floating-point types are currently supported:
\begin{itemize}
    \item 64-bit IEEE double-precision ($\sim 16$ decimal digits),
    \item 80-bit IEEE extended-precision ($\sim 19$ decimal digits),
    \item 128-bit IEEE quadruple-precision ($\sim 34$ decimal digits).
\end{itemize}
The double-precision format is available on all compilers and hardware platforms as the C++
\texttt{double} type. The 80-bit extended precision format is available on x86 processors, and it can be
used on some compiler/operating system combinations via the C++ \texttt{long double} type. The
quadruple-precision format is available on most 64-bit platforms when using the GCC or Clang compilers,
where it is available in C++ via the nonstandard \texttt{\_\_float128} type.

While the 64-bit and 80-bit floating-point types are implemented in hardware, the 128-bit floating-point
type is typically emulated in software\footnote{Note, however, that recent versions of the PowerPC architecture
feature quadruple-precision arithmetic implemented in hardware.}, and thus it is usually
$\gtrsim 2$ orders of magnitude slower than
the other floating-point types.

\subsection{Batch mode}
\label{subsec:batch}
When performing numerical experiments, the need to integrate an ODE system with a range
of different initial conditions often arises (e.g., in parametric studies). This mode of usage
provides straightforward opportunities for parallelisation at a coarse level on multicore CPUs
(i.e., by running multiple independent integrations at the same time).
Modern CPUs feature another, finer-grain level of parallelisation in the form of SIMD (single instruction, multiple data) vector instructions,
which allow to perform mathematical operations on small floating-point vectors at a cost similar to scalar
operations (thus providing a potential increase in floating-point
throughput that scales linearly with the SIMD vector size).

In order to fully exploit the SIMD capabilities of modern CPUs, our implementation
supports a \emph{batch mode} option. In an ODE integrator created in batch mode, all scalar
quantities (i.e., initial conditions, state variables, time coordinate, integration timestep, etc.) are replaced
by SIMD vectors, so that at every timestep the integrator is effectively propagating multiple
trajectories at the same time, each with its own adaptive integration timestep.
As we will show in \S\ref{subsec:batch_benchs}, batch mode can lead to a substantial floating-point throughput increase.

Note however that
batch mode is effective only when operating in double precision, since no CPU at this time supports SIMD instructions
on extended-precision datatypes. Additionally, because modern CPUs do not provide SIMD instructions
for the computation of elementary functions (apart from \texttt{sqrt()}), in batch mode
we rely on \texttt{SLEEF} (SIMD library for evaluating elementary functions) by \citet{shibata2019sleef} to provide
a fully vectorized integration loop.

\subsection{Just-in-time compilation and compact mode}
\label{sec:jitc}
As we mentioned earlier, and contrary to existing implementations of Taylor methods, our implementation does
not output a set of source files to be compiled separately. Instead, we use the
LLVM compiler infrastructure \citep{lattner2004llvm} to synthesize an ODE integrator at runtime
via a process of just-in-time compilation.

LLVM is designed around a language-independent intermediate representation (IR), a
portable, high-level assembly-like language that can be optimized with a variety of transformations.
We generate the integrator code (which includes the automatic differentiation rules,
the adaptive timestep deduction and the state propagation via polynomial evaluation) in the LLVM IR language,
and then compile it on-the-fly using LLVM's just-in-time compilation capabilities.
Because the
binary code of the integrator is generated at runtime, our implementation is able to aggressively optimise
the code for the CPU on which the code is running (e.g., by taking advantage of the specific SIMD
instructions available on the CPU).
The final result
of this process is a function pointer to a timestepping function that can be used in normal C++ code.

By default, we aim at maximising the integrator performance above everything else. Specifically,
our implementation will, by default, produce a timestepping function without branching instructions in which all loops
and summations have been manually unrolled.
This approach, while producing highly efficient code, leads however
to long compilation times and binary size bloat for large ODE systems. Thus, we also provide
a \emph{compact mode} option in which
\begin{itemize}
    \item summations are implemented as \texttt{for} loops (instead of being fully unrolled), and
    \item repeated patterns in the decomposition of an ODE system into elementary subexpressions are recognized
    and compressed into \texttt{for} loops.
\end{itemize}
This approach
drastically curbs compilation times, at the price of a performance degradation that, in our numerical experiments,
we have measured roughly as a factor of $\lesssim 2\times$ (see also the results in \S\ref{subsec:st_nbody}).
As we will show in \S\ref{subsec:st_nbody} and \S\ref{subsec:mascons},
the pattern recognition capabilities of our implementation
render it usable on ODE systems with a large number of terms/equations.

\begin{figure}
  \centering
  \includegraphics[width=\columnwidth]{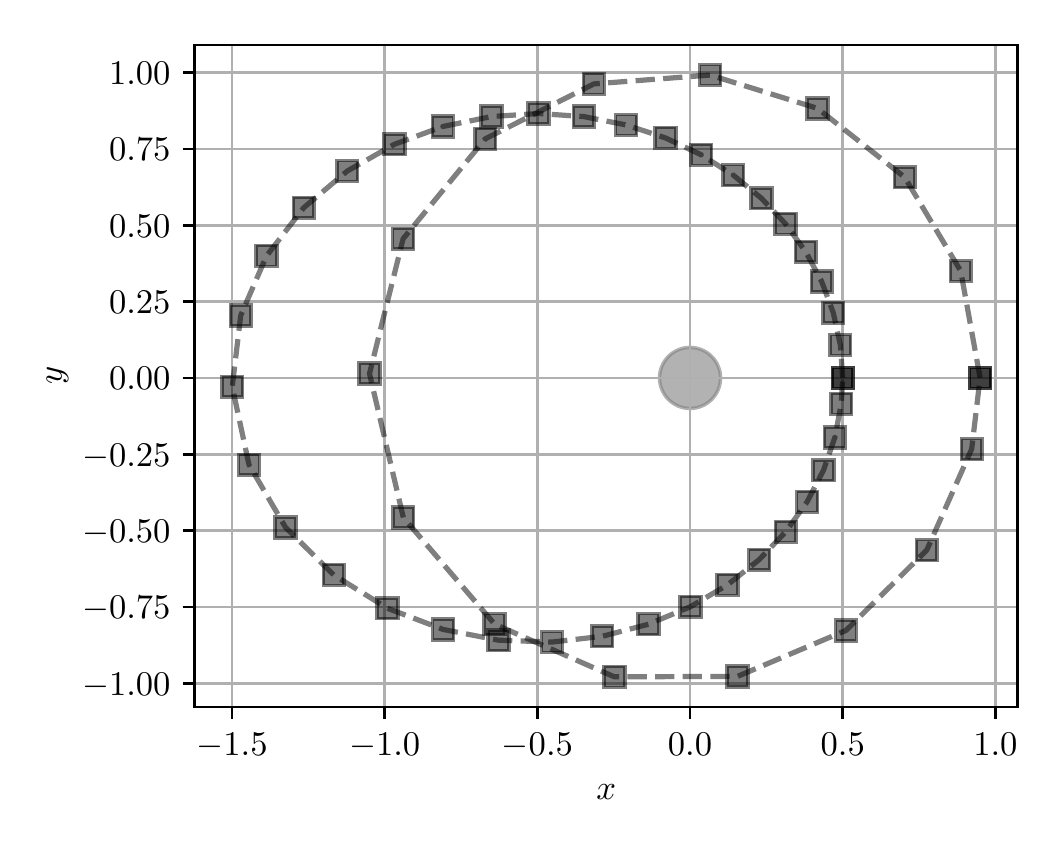}
  \caption{Propagation of two Keplerian orbits (with eccentricities 0.05 and 0.5) with our Taylor integrator. The error tolerance
  is set to $2.2\times 10^{-16}$, the centre of force is in the origin and the pericentre lies on the $x$ axis.
  The position at the end of each integration timestep is marked with a square. With these settings,
  our integrator requires $\sim 16$ steps for the integration of a full low-eccentricity orbit.}
  \label{fig:ob_00}
\end{figure}

\subsection{High accuracy mode}
\label{subsec:high_acc_mode}
When dealing with long-term integrations of dynamical systems, in addition to ensure that
the integration error is kept below the desired threshold on a step-by-step basis, it is also important
to minimise any numerical bias in the integration error. If the integration error has the statistical
properties of a random distribution, then it will accumulate over the integration time $t$ proportionally to $\sqrt{t}$ (i.e.,
as a one-dimensional random walk). This result is usually referred to as Brouwer's law \citep{brouwer1937accumulation}.
Because IEEE floating-point arithmetic guarantees that, for random operands,
the error is indeed randomly distributed, Brouwer's law represents the optimal outcome
achievable on a real computer for a numerical integrator. Deviations from Brouwer's law (e.g., a linear increase over time
of the accumulated error) indicate a source of bias in the integrator.

Bias in an integrator may originate from a variety of sources, including the integration scheme,
specific implementation details and even compilation flags. We refer to \S3 of \citet{rein2015ias15} for an excellent,
in-depth discussion of possible sources of biases and systematic errors in numerical integrators. From a practical
standpoint, a simple way of reducing bias is to decrease the error tolerance in the integration scheme
below machine precision. For instance, if we are operating in double-precision arithmetic, machine precision
is $\sim 10^{-16}$. If we are able to implement the integration loop in extended precision,
then any bias in the integrator will be hidden behind the noise of floating-point arithmetic (which follows
Brouwer's law), at least for a certain amount of time.

From the point of view of the integration scheme, in Taylor integrators it is straightforward to operate
in extended precision: we just need to increase the Taylor order $p$ above the level suggested by eq. \eqref{eq:opt_order_00}
(or, equivalently, decrease the error tolerance below machine precision).
An alternative would be to reduce the timestep $h$ (this is the approach taken by \citet{rein2015ias15}), but,
as shown in \S\ref{sec:comp_high_acc}, increasing $p$ is a more efficient way of achieving the same goal.

In order to reduce bias, we also need to extend the precision of floating-point arithmetic. A way of achieving
this goal would be to use double-length arithmetic \citep{dekker1971floating} within the timestepper code. However,
as pointed out in \citet{abad2012algorithm}, \citet{rodriguez2012reducing} and \citet{rein2015ias15}, a faster and more practical way of reducing numerical
errors is to control catastrophic cancellations in summations via compensated algorithms\footnote{Note that
compensated summation algorithms are essentially approximate versions of double-length arithmetic, as both methods
are based on error-free transformation techniques \citep{ogita2005accurate, graillat2008compensated}.}
\citep{kahan1965pracniques,neumaier1973,higham1993accuracy,graillat2008compensated}.
To account for the effects discussed above and ensure a minimal bias we use pairwise summations (see \citet{higham1993accuracy} for details on the algorithm) in the implementation of all automatic differentiation formul\ae{} whenever possible. Additionally, the evaluation of the Taylor polynomials \eqref{eq:taylor_poly_00} may be performed, instead of via Horner's method, via a compensated summation. Since the compensated summation introduces a measurable computational burden we make its use optional via the activation of a \emph{high accuracy mode} in \heyoka{}.

As we will show in \S\ref{subsec:nbody},
high Taylor orders and these error-reduction techniques allow our implementation to respect Brouwer's law for (at least) one billion dynamical timescales.

\section{Tests}
\label{sec:tests}
In this section we will present a variety of tests of astrodynamical interest aimed at evaluating the performance of our Taylor integrator.
Our objective is to assess both the runtime performance and the capability to preserve dynamical invariants
to the desired level of precision.
Unless otherwise specified, all measurements were conducted on a AMD Ryzen 3950X CPU in a 64-bit Linux environment.

\subsection{Central force problem}
In order to show the qualitative behaviour of our Taylor integrator, as a preliminary test we integrate
two Keplerian one-body orbits with different eccentricities (0.05 and 0.5).
The integration is performed in double-precision arithmetic
with an error tolerance of $2.2\times 10^{-16}$. According to eq. \eqref{eq:opt_order_00}, this tolerance corresponds
to a Taylor order $p=20$.

The results are displayed in Figure \ref{fig:ob_00}, where we mark with a square the position of the test
particle at the end of each adaptive timestep. The integration starts at the pericentre, which is located on the $x$ axis.
The centre of force is located in the origin.
As expected, the adaptive timestep deduction scheme described in \S\ref{subsec:timestep} yields longer
timesteps close to the apocentre and shorter timesteps near the pericentre. Thanks to the high Taylor order $p=20$,
in the low-eccentricity orbit the integrator is able to maintain machine precision while using
$\sim 16$ steps per orbit. For comparison, the 15-th
order \texttt{IAS15} integrator requires about 100 steps per orbit in order to achieve the same precision
\citep[\S3.7]{rein2015ias15}. In these simple examples, the relative energy error after one orbit is $\sim 10^{-16}$,
and at the end of one orbital period the test particle is back at the pericentre with an error of $\sim 10^{-15}$.

\subsection{Short-term N-body integrations}
\label{subsec:st_nbody}
As a second test aimed to evaluate the runtime performance of our integrator, we set up an ODE system representing a
planetary system with a Sun-like star and a variable number of planets on (initially) circular orbits.
The mass and initial semi-major axis of the $n$-th planet in the system are set, respectively, to
$1/{n^2}$ Earth masses and $n$ astronomical units. Because our focus in this test is performance assessment,
we set a short integration time of $1000$ years.
For this test, we operate with double-precision arithmetic, we set the tolerance
to $10^{-18}$ and we enable high accuracy mode (see \S\ref{subsec:high_acc_mode}). This is the same setup
adopted in \S\ref{subsec:nbody}, which, as we will show later, allows to fulfill Brouwer's law
in long-term integrations.

\begin{figure}
  \centering
  \includegraphics[width=\columnwidth]{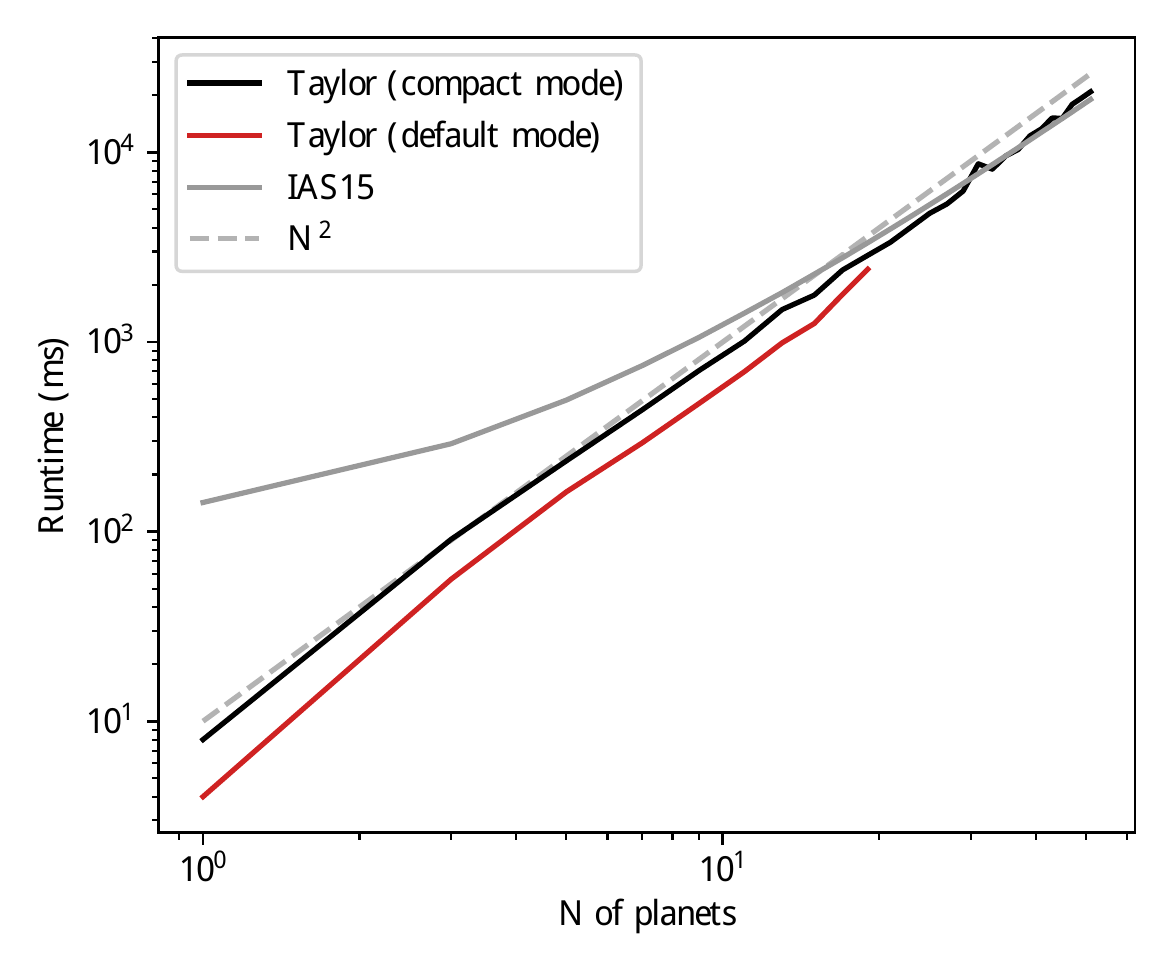}
  \caption{Performance comparison between our Taylor integrator and \texttt{IAS15} \citep{rein2015ias15} in short-term N-body integrations. The
  system consists of a central Sun-like star surrounded by an increasing number of planets (shown on the $x$ axis). The runtime is given
  in milliseconds ($y$ axis), and the total integration time is $1000$ years. Our integrator is tested both in default and compact mode (see \S\ref{sec:jitc}).}
  \label{fig:ss_maker_00}
\end{figure}

For this test, we compare our implementation to \texttt{IAS15} \citep{rein2015ias15}.
The test for our integrator was run both in default and compact mode (see \S\ref{sec:jitc}).
The results of the comparison are visualized in Figure \ref{fig:ss_maker_00}. The plot clearly shows that
the just-in-time compilation process of our implementation is able to produce a highly optimised integrator
when the number of bodies is low. In the extreme case of $N=1$ (i.e., a two-body problem), our integrator in default
mode is $\sim 35$ times faster than \texttt{IAS15}. As the number of bodies increases, the computational
load is increasingly dominated by the evaluation of the gravitational forces between the bodies.
Eventually, around $N\sim 37$, \texttt{IAS15} matches the performance of our implementation, and for higher
values of $N$ both integrators settle into a quadratic dependence on $N$ (see the grey dashed line in Figure \ref{fig:ss_maker_00}).
As mentioned earlier,
compact mode (see \S\ref{sec:jitc}) results in a performance penalty of roughly a factor
of 2 with respect to the default code generation mode. The compilation time for the default code generation
mode, however, scales poorly with $N$, reaching a value of $\sim \SI{900}{\second}$ for $N=19$, and it is thus
not practical for higher values of $N$. By contrast, the compilation time in compact mode for $N=19$
is $\sim \SI{0.19}{\second}$.

We need to emphasise that large N-body systems are particularly challenging for our Taylor integrator,
especially compared to a special-purpose N-body integrator such as \texttt{IAS15}.
Some of the challenges are intrinsic to the Taylor integration method, while other challenges derive from
the general-purpose character of our implementation. We refer to \S\ref{sec:limitations} for an in-depth
discussion of various performance issues and potential improvements that can lead to more efficient
special-purpose Taylor integrators.

\subsection{Long-term N-body integrations}
\label{subsec:nbody}
In this test, we focus on the behaviour of our Taylor integrator in a long-term integration of the outer Solar System.

Long-term integrations of the Solar System are often undertaken with symplectic integrators,
which guarantee (from the point of view of the integration scheme) the conservation of dynamical invariants such
as the total energy of the system \citep{de1956methods,ruth1983canonical}. Symplectic integrators widely used in Celestial
Mechanics include the Wisdom-Holman (WH) integrator \citep{wisdom1991symplectic,wisdom2018dynamical}
and the mixed-variable integrator of \texttt{MERCURY} \citep{1997DPS....29.2706C}. More recently, high-order symplectic integration schemes
have been proposed and successfully employed to investigate
the dynamical evolution of the Solar System over hundreds of millions of years
\citep{laskar2001high,laskar2009existence,blanes2013new,rein2019high}. As pointed out in \citet{rein2015ias15},
however, symplectic integrators also suffer from a variety of drawbacks, including:
\begin{itemize}
    \item difficulties in making the timestep  adaptive while  keeping  the symplecticity,
    \item reliance on a hierarchical setup of the dynamical system (i.e., one body has to be identified
    as the ``star''),
    \item difficulties in handling non-conservative forces (e.g., solar radiation pressure, gas drag, etc.).
\end{itemize}
Additionally, \citet{rein2015ias15} show that their non-symplectic \texttt{IAS15} integrator (an implicit
Runge-Kutta method based on Gau{\ss}-Radau quadrature) is able to conserve energy over a billion dynamical
timescales at least as well as any symplectic integrator they tested (see also the results in \citet{wisdom2018dynamical}
for a comparison between \texttt{IAS15} and a modern implementation of the WH scheme).

To the best of our knowledge, the question of whether Taylor methods are suitable for long-term integrations
of the Solar System is unanswered in the literature. Previous applications of Taylor integrators to problems of
celestial mechanics \citep{jorba2005,barrio2005performance,abad2012algorithm,rodriguez2012reducing} demonstrate good energy conservation
properties over short timescales, but the long-term behaviour is not addressed. \citet{jorba2005} in particular
pose the question of whether ``Taylor methods can compete with symplectic integrators in the preservation of the geometrical structure of
the phase space of a Hamiltonian system'', without however providing an answer.

For this test we will be integrating the outer Solar System (that is, the Sun, Jupiter, Saturn, Uranus, Neptune and Pluto) as a full
6-body problem for $10^{10}$ years (that is, roughly $10^9$ Jupiter orbits, or dynamical timescales). The initial conditions are taken
from \citet{applegate1986outer}. In order to add statistical weight to our test, we will run 20 distinct integrations
in which the initial conditions are slightly
and randomly altered with respect to the values reported in \citet{applegate1986outer}.
The integration is carried out in double precision, with a tolerance of $10^{-18}$ (corresponding to a Taylor order $p=22$)
and in high accuracy mode (see \S\ref{subsec:high_acc_mode}). Note that,
because this experiment involves a numerical integration over many Lyapunov times,
the high numerical accuracy does not necessarily confer a special accuracy on the study of the outer Solar System
\citep{zwart2014minimal,boekholt2015reliability}.
Indeed, the intent of this test is specifically to study the energy conservation properties of \heyoka{} in long-term
integrations.

\begin{figure}
  \centering
  \includegraphics[width=1\columnwidth]{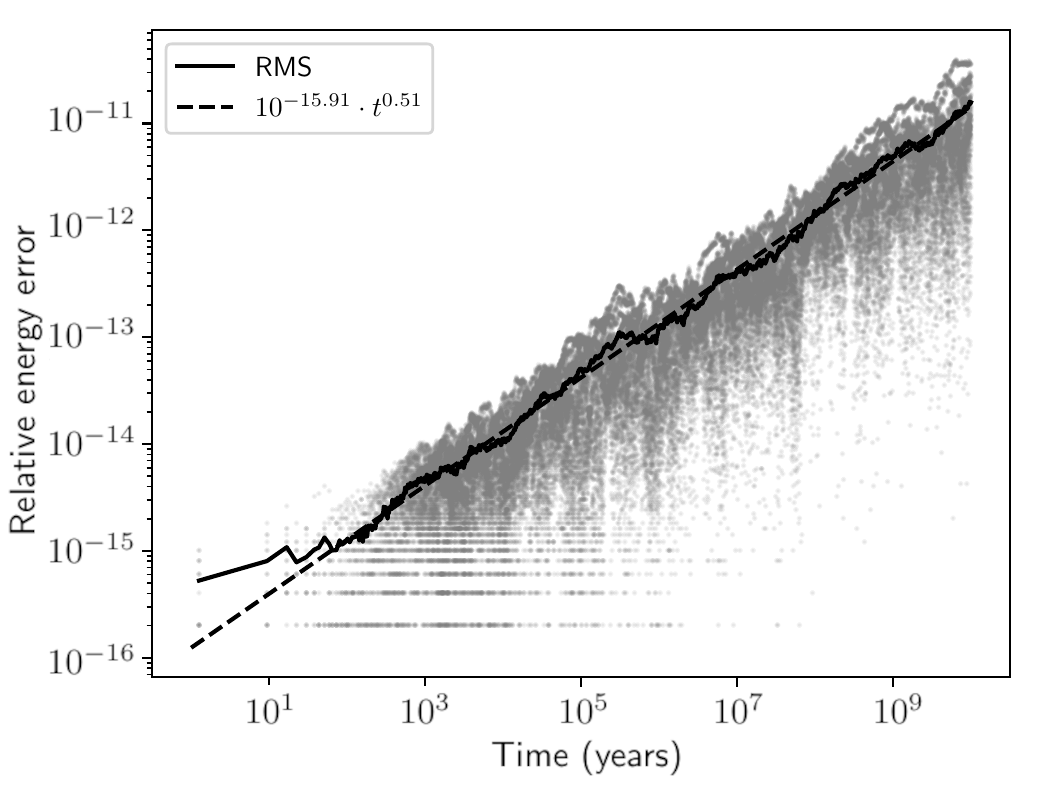}
  \caption{Evolution in time of the relative energy error in long-term integrations of the outer Solar System using our Taylor method. The grey datapoints correspond to 20 distinct simulations in which
  the initial conditions are slightly and randomly perturbed with respect to the values published in \citet{applegate1986outer}. The solid black line represents
  the root mean square (RMS) of the relative energy error across the 20 simulations. The dashed line is the linear fit (in logarithmic space) of the RMS of the relative energy error starting
  from $t=100\,\textnormal{y}$.}
  \label{fig:lt_00}
\end{figure}

The results are displayed in Figure \ref{fig:lt_00}, where we plot the relative energy error over time for each of the 20 simulations, the RMS
of the energy error across the 20 simulations and its linear fit in logarithmic space.
The linear fit clearly indicates that the energy error is, on average, growing in time as $\sim \sqrt{t}$. Thus, our Taylor integrator
obeys Brouwer's law up to (at least) one billion dynamical timescales.

From the performance point of view, our integrator takes $\sim \SI{14.9}{\second}$ to integrate the outer Solar System for $10^6\,\textnormal{y}$.
For comparison, in the same setup \texttt{IAS15} (which, we recall, compares favorably to various symplectic integrators, both in accuracy and performance,
as reported in \citet{rein2015ias15} and \citet{wisdom2018dynamical}) takes $\sim \SI{58.7}{\second}$. These timings are consistent with those
reported in \S\ref{subsec:st_nbody}. These results indicate that Taylor integrators are
a suitable tool for studying the long-term dynamics of the Solar System, and that they are competitive (in terms of both performance and accuracy)
with state-of-the-art symplectic and non-symplectic integrators.

The choice of a tolerance of $10^{-18}$ is motivated, as explained
in \S\ref{subsec:high_acc_mode}, by the necessity of lowering the error of the integration scheme below machine precision
in order to prevent bias in long-term integrations. We determined that the specific value $10^{-18}$ is
low enough to satisfy Brouwer's law for the conservation of energy over billions of dynamical timescales, as shown by the numerical
experiments in this section and in \S\ref{subsec:trappist1}. A lower tolerance value may be necessary to prevent bias
over even longer timescales.

Our results confirm and extend the findings of \citet{rodriguez2012reducing}, who conclude, via analytical arguments
and numerical experiments, that the combined use of error tolerances below machine precision and compensated summation
techniques is sufficient to suppress long-term bias, thus allowing to achieve Brouwer's law
(for the conservation not only of the energy, but also of all the first integrals of the system),
albeit for timescales much shorter than those considered here.
One peculiar finding from our work, which is also confirmed by \citet{rein2015ias15} and
\citet{hernandez2021enckehh}, is that the use of adaptive timestepping helps reducing the accumulation of bias over long-term integrations.
We have verified that forcing a constant timestep in our integrator results in a much faster bias accumulation with respect to the adaptive timestepping strategy.
Interestingly, adding a component of random noise to the fixed timestep helps reducing the accumulation of bias.

Regarding the comparison with \texttt{IAS15}, we also need to point out that both our tests (which were
performed with the latest version of \texttt{REBOUND} \citep{rein2012rebound})
and the results reported in \citet{rein2015ias15} indicate that \texttt{IAS15} is about half a decimal digit more precise than our Taylor integrator.
That is, while for both \texttt{IAS15} and our Taylor integrator the energy error grows as $\sqrt{t}$ in long-term integrations, the
constant factor in front of $\sqrt{t}$ is slightly lower for \texttt{IAS15}. This difference persists even when setting the tolerance
to much lower values (e.g., $10^{-24}$) and it is also unaffected by high-accuracy mode. Because of this, we believe that
higher accuracy of \texttt{IAS15} may be related to the specific way in which the AD formul\ae{} are implemented in \heyoka{}
(although we have not
been able to definitively confirm this hypothesis).

\subsection{Kirkwood gaps}
\label{subsec:kirkwood}
In this test our goal is to simulate the formation of the Kirkwood gaps in the asteroid belt. The Kirkwood gaps are dips in the distribution of
the semi-major axes of the orbits of main-belt asteroids in correspondence of mean-motion resonances with Jupiter. Although the connection
between the location of the gaps and the mean-motion resonances has been known since the late 19th century, a detailed explanation
of the dynamical mechanisms that lead to the formation of the gaps has been achieved only recently. The Kirkwood
gaps are now understood to originate from the overlapping of secular resonances within the mean-motion
resonances, which excite the asteroids' eccentricities to high values. Highly eccentric asteroids are then removed from the resonances
following close encounters with the planets, or by plunging into the Sun. See \citet[chapter 11]{morby_mcm} (and references
therein) for a detailed analysis of secular dynamics inside mean motion resonances.

Simulations of the formation of the Kirkwood gaps are usually undertaken with the help of mixed-variable symplectic integrators such as \texttt{MERCURY} \citet{1997DPS....29.2706C}, which provide great performance while guaranteeing the conservation of the total energy. Note however that simulations in a high-precision setting such as those performed here guarantee the conservation of all first integrals (not only of the energy), and thus can provide an important validation.

For this numerical experiment we construct a dynamical system consisting of:
\begin{itemize}
    \item the Sun, Jupiter and Saturn as a full 3-body problem, with initial conditions taken from \citet{applegate1986outer},
    \item 25600 asteroids, modelled as point masses, placed on randomly-generated low-eccentricity and low-inclination orbits
    in the asteroid belt,
    \item Mars, the Earth and Venus modelled as massive bodies on circular orbits around the Sun. In this simplified model, the three inner planets
    are a time-dependent perturbation on the dynamics of the asteroids, but they do not interact with
    the Sun or with the giant planets.
\end{itemize}
We integrate the system for a total of $5\,\textnormal{My}$, and we track the evolution of the asteroids' orbital elements over time.
As in \S\ref{subsec:nbody}, we operate in double precision with a tolerance of $10^{-18}$, and we activate high accuracy
mode. For this experiment we also enable batch mode (see \S\ref{subsec:batch} and \S\ref{subsec:batch_benchs}), which allows
us to increase the number of simulated asteroids by a factor of 4 at the price of a slight increase in the overall latency
of the simulation.

\begin{figure*}
  \centering
   \includegraphics[width=.6\textwidth]{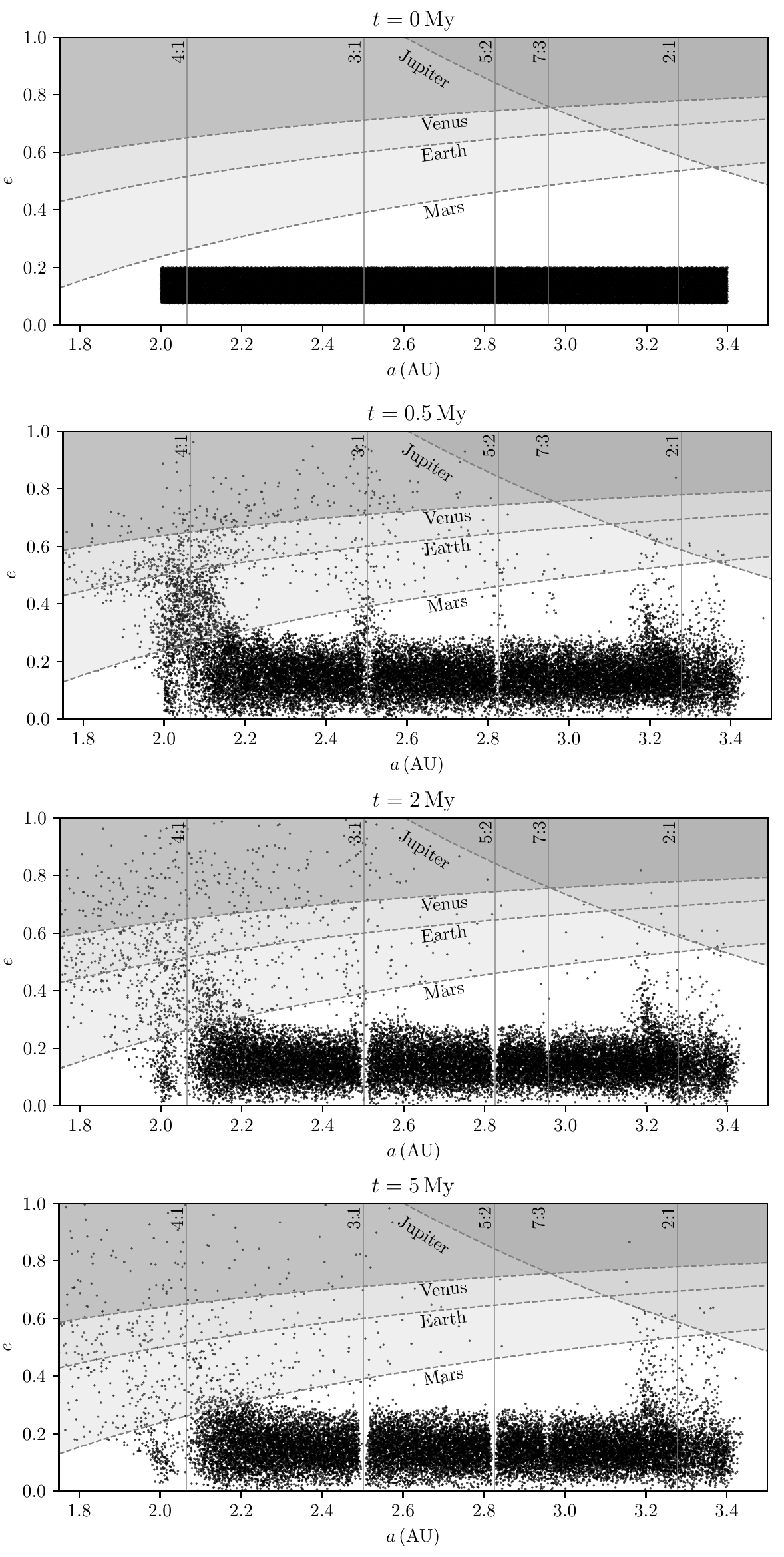}
  \caption{Simulation of the formation of the Kirkwood gaps in the asteroid belt (see \S\ref{subsec:kirkwood}). At $t=0$, 25600 asteroids
  are placed on randomly-generated, low-eccentricity and low-inclination orbits in the asteroid belt (top panel). The orbits of the asteroids
  are perturbed by Jupiter, Saturn, Mars, the Earth and Venus. At $t=0.5\,\textnormal{My}$ (second panel), gaps start to be visible in correspondence of the main
  mean-motion resonances in the belt, where the asteroids' eccentricities increase to planet-crossing values. As the simulation progresses,
  the gaps become more defined (third and fourth panels). In the plots, the locations of the main mean-motion resonances are
  marked with vertical lines. The dashed lines represent eccentricity values above which the asteroids can experience close encounters with the planets.}
  \label{fig:kirkwood_global}
\end{figure*}

\begin{figure}
  \centering
   \includegraphics[width=1\columnwidth]{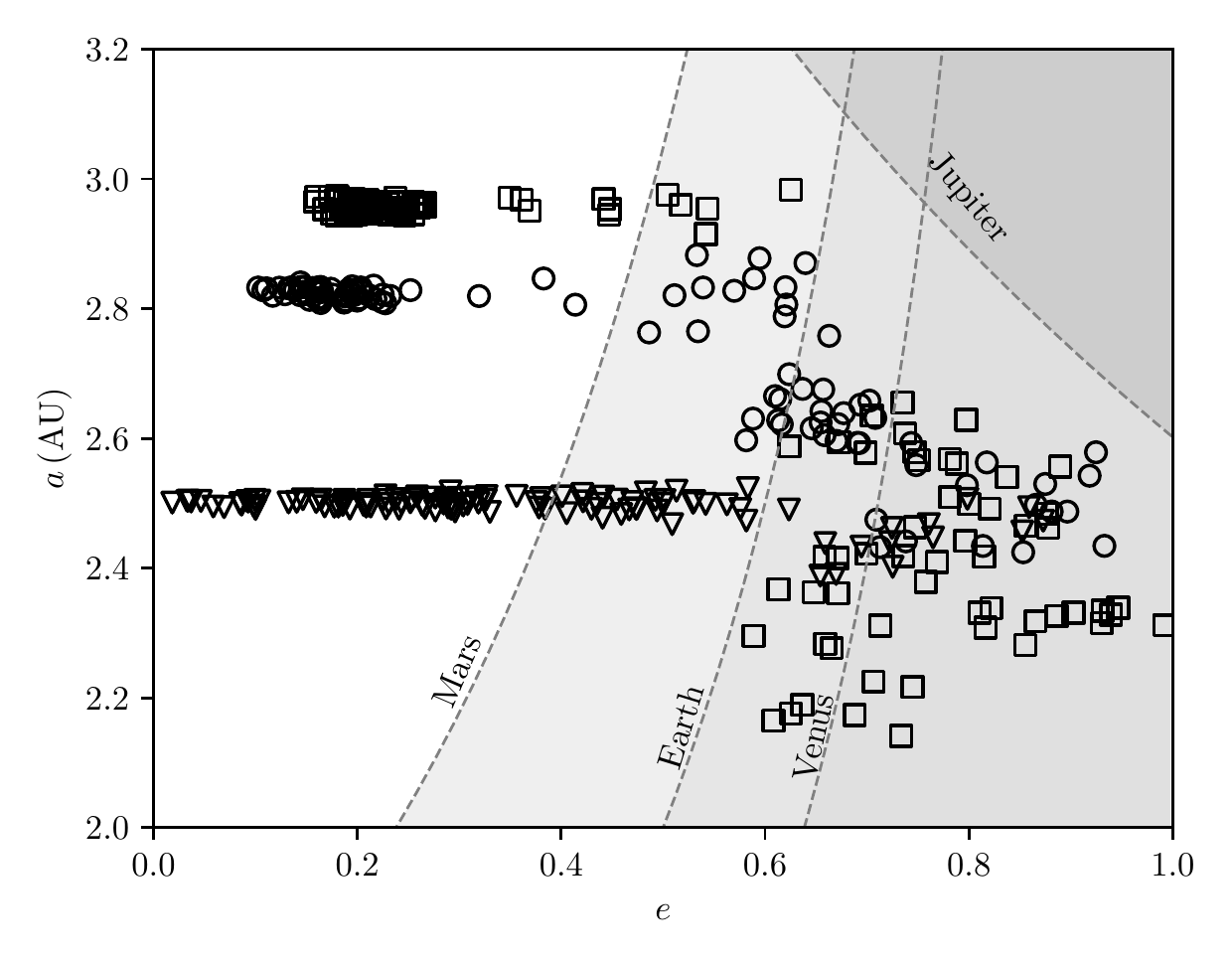}
  \caption{Representative evolutions of three asteroids initially placed inside mean-motion resonances within the asteroid belt. The triangle starts
  in the $3{:}1$ resonance, the circle in the $5{:}2$ resonance and the square in the $7{:}3$ resonance. The eccentricities of the asteroids
  increase over time (while keeping the semi-major axis approximately constant) until the asteroids start crossing the orbits of the planets (dashed lines). Planetary close
  encounters then change the orbital parameters, kicking the asteroids outside the resonances. Close encounters with Jupiter can lead to ejection from the Solar System. In all three cases shown here, the asteroids end up colliding with the Sun by achieving $e \sim 1$.}
  \label{fig:kirkwood_evo}
\end{figure}

The results of the simulation are displayed in Figure \ref{fig:kirkwood_global}, where we plot the semi-major axis and eccentricity
of the asteroids at four different times in the simulation. In the plots, we also mark the locations of the main mean-motion resonances
as vertical lines. The dashed lines represent eccentricity values above which the asteroids can experience close encounters with the planets.

The plots show that the formation of the gaps is well underway at $0.5\,\textnormal{My}$ (top-right panel in Figure \ref{fig:kirkwood_global}),
with asteroids in mean-motion resonances undergoing evolution to high eccentricities and experiencing close encounters with the planets.
At $2\,\textnormal{My}$ (bottom-left panel in Figure \ref{fig:kirkwood_global}) the $4{:}1$, $3{:}1$ and $5{:}2$ gaps are clearly visible,
while the $7{:}3$ gap is beginning to open. Finally, at $5\,\textnormal{My}$ (bottom-right panel in Figure \ref{fig:kirkwood_global})
the gaps have further cleared up and the cloud of low semi-major axis, planet-crossing asteroids on the left side of the
panel has thinned considerably with respect to the $2\,\textnormal{My}$ snapshot as a consequence of planetary encounters. In
Figure \ref{fig:kirkwood_evo}, we show the evolutionary history of three asteroids which are initially placed inside mean
motion resonances and which eventually end up colliding with the Sun.

These outcomes are consistent with the results from \citet{gladman1997dynamical} and \citet{morby_mcm}, which report median asteroid lifetimes
in the $3{:}1$, $5{:}2$ and $7{:}3$ resonances of, respectively, $2\,\textnormal{My}$, $0.6\,\textnormal{My}$ and $19\,\textnormal{My}$. Regarding
the $2{:}1$ resonance, we remark that, although our simulation reproduces the expected eccentricity-pumping behaviour, an integration time of
$5\,\textnormal{My}$ is not enough to deplete the gap, because the $2{:}1$ resonance is characterized by large islands of weakly-chaotic behaviour
in which the diffusion to high eccentricities is much slower than for the other resonances. Specifically, \citet[\S 11.3.2]{morby_mcm} reports
a median lifetime $\gtrsim 1\,\textnormal{Gy}$ for the particles in the weakly-chaotic islands within the $2{:}1$ resonance.

\subsection{Dynamics around mascon models}
\label{subsec:mascons}
Here we consider the orbital dynamics of a spacecraft affected by an irregular gravity field produced by a non spherical central body, e.g., an asteroid, uniformly rotating around its largest principal axis. The body 
gravity is described using a mascon model, that is, by a large number $N$ of points having mass $m_j$. These models are very good at providing a precise description of irregular gravity fields such as those generated by asteroids, with the exception of the surface proximity \citep{park2010estimating} where some loss of precision is unavoidable. 
The resulting equations of motion, written in a reference frame attached to the rotating body, take the form:
\begin{equation}
    \label{eq:eom_mascon}
\ddot {\mathbf r} = -G \sum_{j=0}^N \frac {m_j}{|\mathbf r - \mathbf r_j|^3} (\mathbf r - \mathbf r_j) - 2 \boldsymbol\omega \times \mathbf v - \boldsymbol \omega \times\boldsymbol\omega \times \mathbf r,
\end{equation}
where the $N$ mascons having mass $m_j$ are placed at fixed locations $\mathbf r_j$, $\boldsymbol \omega$ is the asteroid (and the reference frame) angular velocity,
and $\mathbf r, \mathbf v$ represent the spacecraft position and velocity in the rotating frame. 

\begin{figure}
  \centering
   \includegraphics[width=1\columnwidth]{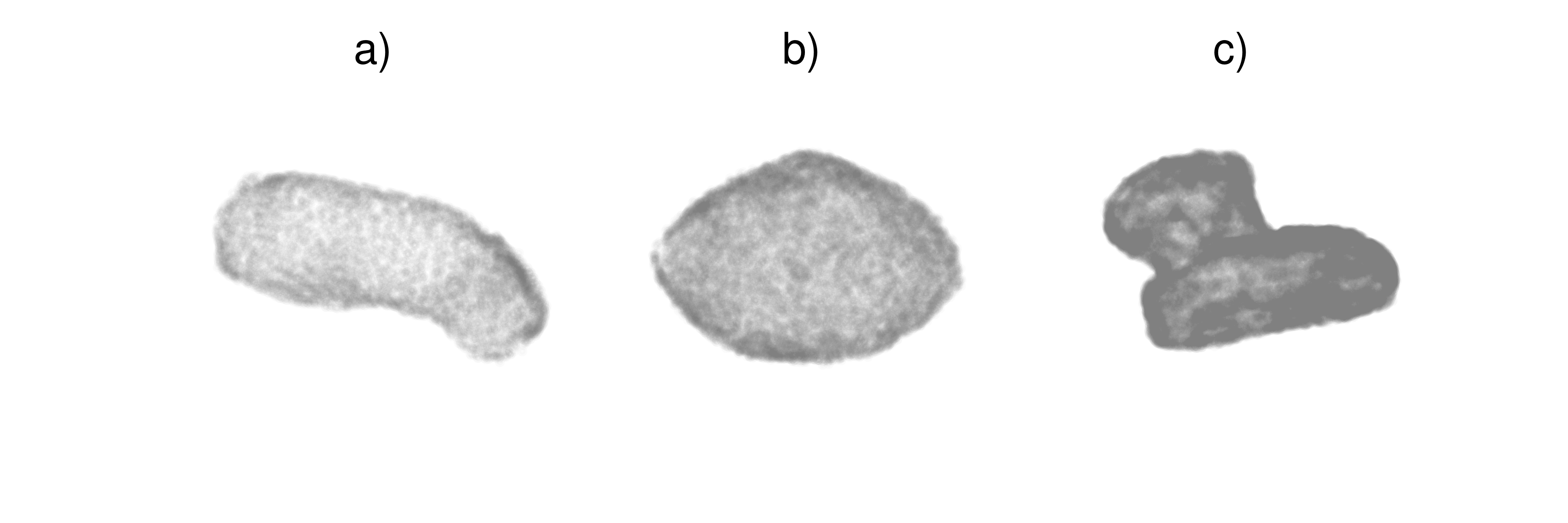}
  \caption{Visualization of the mascon models used. a) Itokawa, b) Bennu, c) Comet 67P/Churyumov-Gerasimenko \label{fig:mascons}.}
\end{figure}

We consider mascon models for the asteroids Itokawa, Bennu and for the comet 67P/Churyumov-Gerasimenko.
The models are created starting from the available three-dimensional surface models of the bodies as distributed by the OSIRIS-REx mission team (for Bennu), the OSIRIS instrument team of ESA's Rosetta mission (for the comet 67P/Churyumov-Gerasimenko) and the AMICA imaging team (for Itokawa).
The triangular surface meshes, after a preliminary resizing step aimed at controlling the final value for $N$, are transformed into a three-dimensional tetrahedral mesh.
One mascon is then placed at the center of each tetrahedron in the mesh and assigned a mass $m_j$ proportional to the tetrahedron volume.
A final transformation is then applied to place the center of mass of the mascon system in the origin of the Cartesian axes and to align its principal axis of inertia (corresponding to the largest moment of inertia) to the z axis.
The final models obtained have an increasing number of mascons: $N=5497$ (Itokawa), $N=12147$ (Bennu) and $N=25033$ (67P/Churyumov-Gerasimenko). The mascon models are visualized in Figure \ref{fig:mascons}.

We compare the results obtained using our Taylor integrator with those of the Runge-Kutta-Fehlberg scheme implemented in the widely used boost C++ odeint library implemented by \cite{ahnert2011odeint}.
In all cases we use nondimensional units setting the Cavendish Constant $G=1$, the total body mass $M=1$ and the following units for the length: $L_{\textnormal{ito}}= \SI{478.26}{\meter}$,
$L_{\textnormal{bennu}}= \SI{416.45}{\meter}$, $L_{\textnormal{67P}}= \SI{2380.7}{\meter}$.The choice of the length units is related to a characteristic dimension of the asteroid
three-dimensional model. The angular velocities of the chosen bodies are set to be, in the chosen units,  $\omega_{\textnormal{ito}}= 0.98$, $\omega_{\textnormal{bennu}}= 1.56$,
$\omega_{\textnormal{67P}}= 0.63$. The initial conditions are $\mathbf r = [3, 0, 0]$ and $\mathbf v = [0, \cos i \sqrt{1 / r}, \sin i \sqrt{1 / r} - \omega]$, $i=45^o$. They correspond to
what would be a Keplerian orbit if all the mass were concentrated in the origin. We set the numerical integration final time to one month thus simulating several
tens of orbits around the bodies.
We record the CPU times as well as the system energy defined as the sum of the kinetic, potential and centrifugal energy:
\begin{equation}
\label{eq:mascon_energy}
J = \frac {v^2}{2} - G\sum_j \frac{m_j}{r_j} - \frac 12 \left(\omega^2r^2 - (\mathbf r \cdot \boldsymbol \omega)^2\right)
\end{equation}
We set for both integrators a target relative accuracy of $\epsilon_r=10^{-14}$ and we make use, also in the RKF7(8) implementation, of the pairwise summation algorithm (see \citet{higham1993accuracy}) as to keep control of the error when performing summations over the $N$ mascons.
\begin{table}
\caption{Performance and accuracy comparison with RKF7(8) over a one month integration time in the mascon models test (see \S\ref{subsec:mascons}).}
\label{tab:rkf78_comparison}
\centering
\begin{tabular}{lcccc}
\toprule
&\multicolumn{2}{c}{\heyoka{}} & \multicolumn{2}{c}{RKF7(8)}\\
&CPU time & Rel. $J$ error & CPU time & Rel. $J$ error \\
\midrule
67P& $\SI{26}{\second}$ & 1.52e-15 & $\SI{53}{\second}$ & 5.41e-15\\
Bennu&$\SI{23}{\second}$ & 5.14e-14 & $\SI{74}{\second}$ &-1.84e-13\\
Itokawa&$\SI{3.6}{\second}$&-3.54e-15 & $\SI{14}{\second}$ & 1.43e-14\\
\bottomrule
\end{tabular}
\end{table}

\begin{figure}
  \centering
   \includegraphics[width=1\columnwidth]{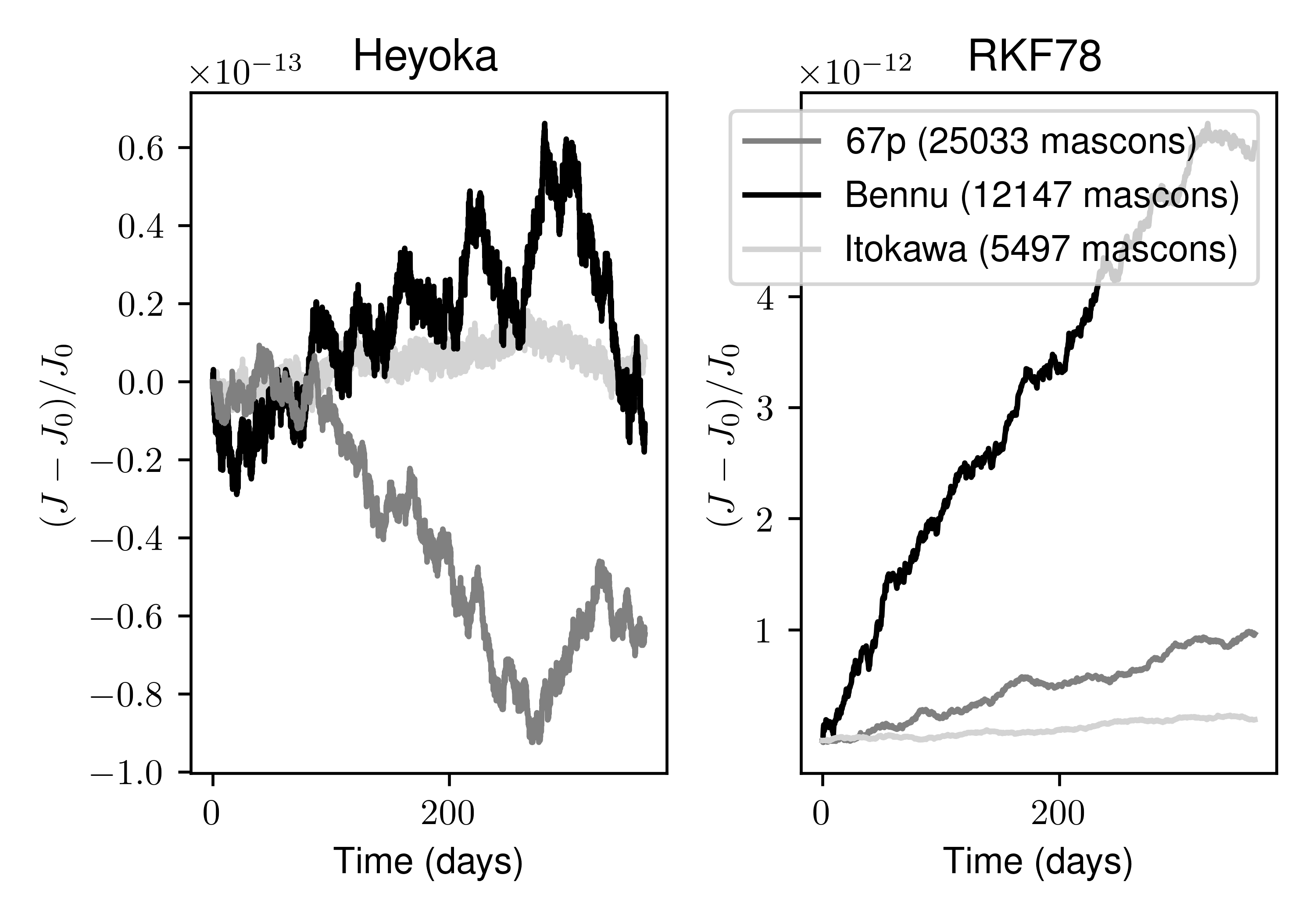}
  \caption{Relative error of the Taylor and the Runga-Kutta-Fehlberg schemes during a whole year of propagation for the three mascon models considered. The error is measured on the relative difference between the starting and the current energy $J$ as defined in eq. \eqref{eq:mascon_energy}.}
   \label{fig:mascons_errors}
\end{figure}

The results are reported in Table \ref{tab:rkf78_comparison} and indicate clearly how our Taylor integrator is able to perform the integration in significantly less time and keeping consistently a better accuracy. It is interesting to study how the error evolves for even longer propagation times. We thus perform again the same experiment, but this time we simulate a whole year of the spacecraft dynamics. The resulting error defined on the energy as $(J - J_0) / J_0$, is visualized in Figure \ref{fig:mascons_errors} and shows, within this relatively long time, a good accuracy ($10^{-14}$) and the absence of long-term error accumulation effects for the Taylor scheme, and confirms its superior behaviour with respect to the Runge-Kutta-Fehlberg scheme that shows instead a clear bias resulting in the error being accumulated steadily.
It is interesting to note how, in previous work, \cite{russell2012global} indicated a weakness of mascon models in their accuracy as observed already during a single orbit propagation. The same work also suggested this issue to be unavoidable and possibly deriving from the large range of magnitudes typically associated to the mascon masses and the consequent inefficiency of having to sum their contributions (even when using pairwise  summation).
In the mascon models considered in our experiments the mass values range from $10^{-3}$ to $10^{-8}$, but we do not observe any loss of precision during the Taylor integration. Our results, for a much longer propagation spanning thousands of orbits, show that an accuracy around $10^{-14}$ is retained throughout the integration period. While our observation may derive from the particular mascon models used, it is more likely linked to the implementation details suggesting, in any case, how the dynamics can be propagated very accurately also when the gravity field is represented by mascon models.

One problem with our Taylor approach is that the numerical integrator needs to be built via the just-in-time compilation process described earlier. In the case of the experiments described 
here and for the sake of completeness, we report the timings of the construction for the three models: $\SI{2.1}{\second}$ for Itokawa, $\SI{4.6}{\second}$ for Bennu and $\SI{9.1}{\second}$ 
for P67. This cost is only paid once, as after the Taylor integrator is constructed it can be reused for any number of different initial conditions.

Note that the system of equations 
\eqref{eq:eom_mascon}, similarly to the N-body case, contains a large summation of similar elements in its right hand side which must be dealt with efficiently when building the Taylor 
integration scheme. The compact mode described in \S\ref{sec:jitc} is just able to do so allowing just-in-time compilation of the Taylor integration also in this case for large values of 
$N$.

\subsection{(99942) Apophis close Earth encounter}
\label{subsec:apophis}
\begin{figure}
  \centering
   \includegraphics[width=1\columnwidth]{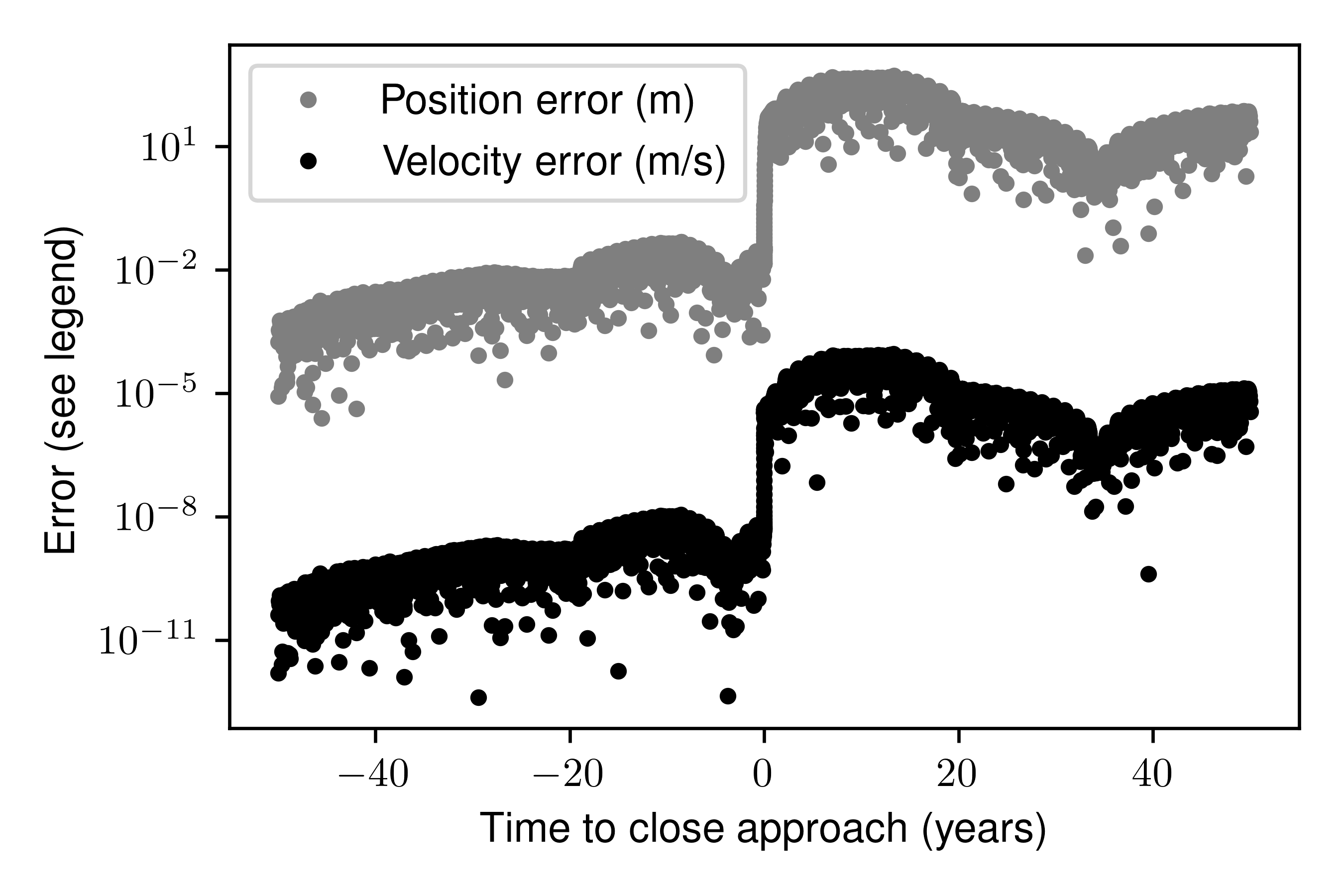}
  \caption{Absolute errors introduced by our Taylor scheme over 100 years in the Apophis close encounter test (see \S\ref{subsec:apophis}).
  The time axis is centered at the epoch of closest encounter with the Earth.\label{fig:apophis_errors}}
\end{figure}

Here we consider the dynamics of the asteroid (99942) Apophis, including its April 2029 close encounter with the Earth. 
For the purpose of this work we consider the Sun, Earth and Apophis system as a full 3-body problem, thus neglecting  contributions such as the gravitational pull of other planets, the Yarkovsky effect and relativistic corrections that play an important role in the actual impact risk analysis of Near Earth Asteroids \citep{farnocchia2013yarkovsky}. We use the JPL Horizon system\footnote{Url: \url{https://http://ssd.jpl.nasa.gov/?horizons} last visited 20
November 2020.} to determine all ephemerides at the epoch of Apophis close Earth encounter 2029-Apr-13 21:42, with respect to the solar system barycenter and in the J2000 reference frame. 
The three body problem is formulated directly in Cartesian coordinates following Cowell's method \citep{bate2020fundamentals}, without making use of any regularization and using directly SI units.
In order to establish a reference ground truth trajectory we use our Taylor scheme to propagate back in time the initial ephemerides using quadruple precision, the high fidelity mode, a tolerance $\epsilon = 1.93\cdot 10^{-34}$ and for a time period of 50 years. 
The resulting asteroid state $\mathbf r_0, \mathbf v_0$ is then propagated, with the same settings, forward in time for 100 years recording the system state and the total energy $J$ at each time step. 
This reference trajectory reproduces, in the employed dynamics, the Apophis close encounter and, in particular its change in semimajor axis from $a>1\,\textnormal{AU}$
(Apollo type) to $a<1\,\textnormal{AU}$ (Atens type). 
A new forward-in-time numerical integration is then performed using our scheme operating in high-fidelity mode, with the tolerance set to $\epsilon =10^{-18}$ and this time using double precision: a set-up that mirrors that used in the previous experiments on short-term propagations of N-body systems (see \S\ref{subsec:nbody}). 
We must mention briefly here, though, that both the decreased tolerance and the high precision mode do not seem to make a great difference in this case since the integration time is relatively short and thus Brouwer's law \citep{brouwer1937accumulation} is easily satisfied in simple double precision.
The results of this second numerical integration are compared to the reference trajectory as to define errors in the asteroid position, velocity and on the total system energy. 
The results, in terms of absolute position and velocity errors, are shown in Figure \ref{fig:apophis_errors}. 
While the close encounter clearly degrades the integration accuracy, Apophis position, some time after the Earth flyby, is still determined within an error significantly smaller than
$\SI{100}{\meter}$.
For comparison, we look into the results reported  in \cite{amato2017accurate}, where a similar Apophis close encounter trajectory is discussed at length.
\begin{table}
\caption{Accuracy and runtime performance comparisons with \citet{amato2017accurate} for the Apophis close encounter test (see \S\ref{subsec:apophis}).}
\label{tab:amato_comparison}
\centering
\begin{tabular}{cccc}
\toprule
& \heyoka{} & XRA15 & LSODAR \\
\midrule
Formulation & Cowell & EDROMO + KS & EDROMO + KS\\
$\delta r$ (m)& $9.23$ & $143.8$ & $5983.9$ \\
$\delta v$ (m/s)& $2.63\cdot10^{-6}$ & $1.31\cdot10^{-6}$ & $5.45\cdot10^{-5}$ \\
Rel. $J$ error & $5.86\cdot10^{-15}$ & $4.81\cdot10^{-16}$ & $1.01\cdot10^{-15}$ \\
CPU times (s) & 0.005 & 0.159& 0.134\\
\bottomrule
\end{tabular}
\end{table}
A special set of orbital elements based on \texttt{DROMO} \citep{urrutxua2016dromo} and of the Kustaanheimo-Stiefel (KS) regularization \citep{kustaanheimo1965perturbation}
is reported to allow for the best precision on the same, 100 year long, numerical propagation centered around the Apophis close encounter to the Earth.
Interestingly, two different numerical integration schemes are considered, \texttt{XRA15} and \texttt{LSODAR}, both based on the Everhart-Radau numerical scheme
\citep{everhart1985efficient}
improved with the adaptive step size strategy proposed by \citet{rein2015ias15}, and thus resulting in numerical schemes similar to \texttt{IAS15}\footnote{One difference being the absence of a compensated summation mechanism that, for these short integration times, does not make a difference.}.
In Table \ref{tab:amato_comparison} we summarize the comparison between our Taylor scheme and the results reported in \citet{amato2017accurate}.
Interestingly, our approach is able to reach better or equivalent precision even if based on the straight forward Cowell's method. This suggests that, for this specific case, the use of \texttt{EDROMO} orbital elements, KS regularization and trajectory splitting methods is not necessary to achieve satisfactory results. The computational time of our approach is also significantly better than that reported by \citet{amato2017accurate} on their approach. While it is always unfair to compare CPU times obtained in heterogeneous setups, we note how this observation is compatible with what reported in Figure \ref{fig:ss_maker_00}, considering that in this case the number of planets is $N=3$ and the numerical techniques we compare to are both very similar to \texttt{IAS15} and thus a similar difference in computational times is also to be expected.

\subsection{TRAPPIST-1 Stability}
\label{subsec:trappist1}
In this test we use our Taylor integration scheme to simulate the dynamics and long term-stability of multiplanetary systems. 
We consider the case of the TRAPPIST-1 system \citep{gillon2017seven}.
TRAPPIST-1 is a system made of seven Earth-sized planets, orbiting closely around a small ultra-cool dwarf with periods forming a near-resonant chain. 
The system is of particular interest as it is considered as a candidate to host extraterrestrial life \citep{lingam2017enhanced} and
it is relatively close to our solar system ($\sim$ 40 light years). 
Only recently, after drastically reducing the uncertainties on the system identification, \citet{agol2020refining} were able to
simulate a 10 My long period of stability starting from a large number of posterior estimates of the initial system state.
Earlier simulations performed by \citet{gillon2017seven} and \citet{grimm2018nature} were, instead, based on higher uncertainties, and found a large amount of unstable configurations, which led to 
suggest that the TRAPPIST-1 system could be unstable over relatively short timescales.
This, in turn, led to conjecture a strong tidal stabilizing effect due to the star proximity \citep{gillon2017seven} as well as an important 
planetary migration \citep{tamayo2017convergent} and the presence of an undiscovered planet stabilizing the system. 
All numerical simulations on the long term stability of the TRAPPIST-1 system, so far, were done using symplectic numerical integrators including \cite{wisdom1991symplectic}, while keeping the relative error on the energy at around $\epsilon = 10^{-7}$. 
For this numerical experiment we integrate the N-body dynamics of the TRAPPIST-1 system aiming at replicating  the results found by \citet{agol2020refining} in a high accuracy setting.

\begin{figure}
  \centering
   \includegraphics[width=0.98\columnwidth]{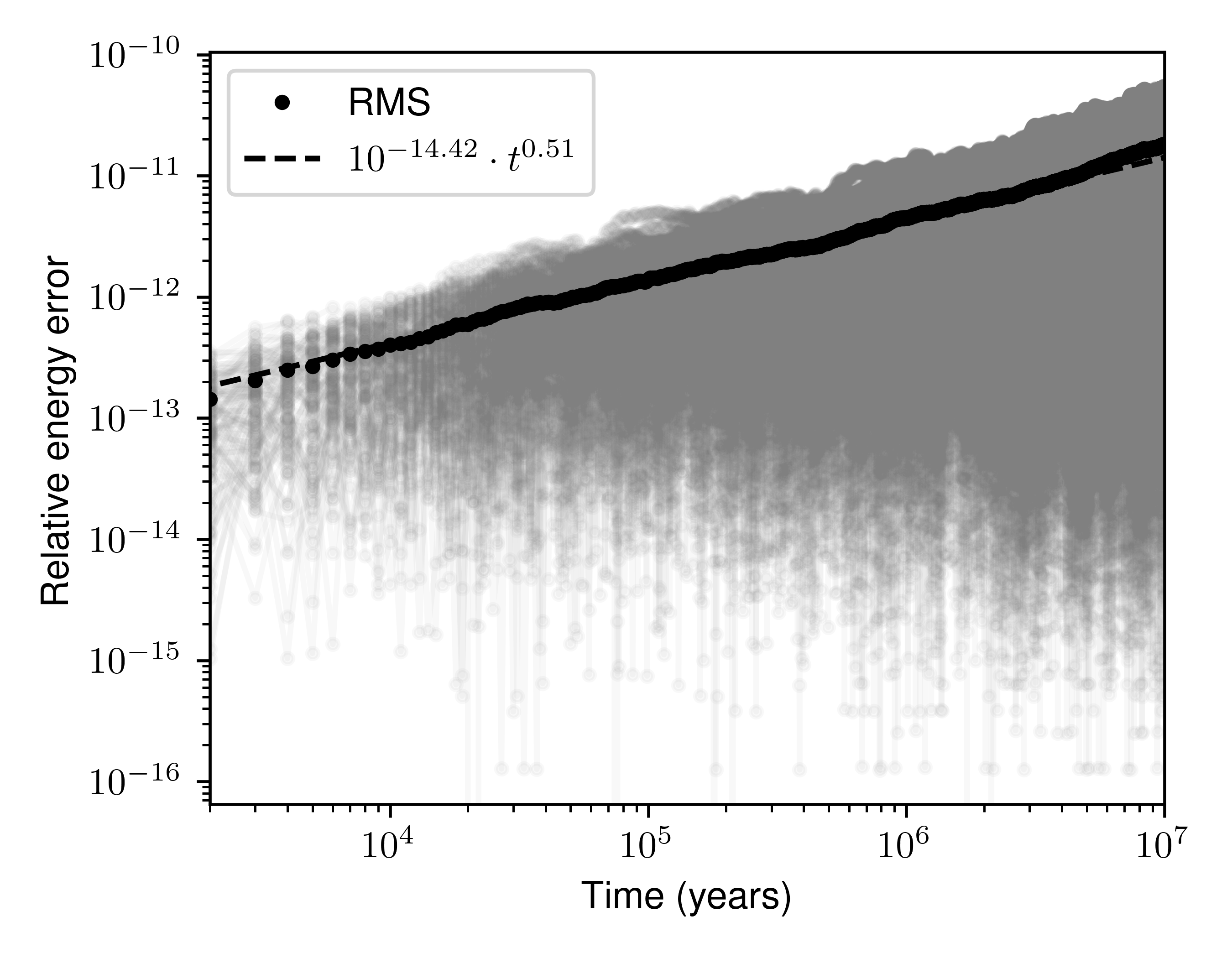}
  \caption{Evolution in time of the relative energy error in long-term integrations of the TRAPPIST-1 system using our Taylor method. The grey datapoints correspond to 128 distinct simulations in which
  the initial conditions are sampled from aposterior probabilitiesa as in \citet{agol2020refining}. The solid black line represents
  the root mean square (RMS) of the relative energy error across the 128 simulations. The dashed line is the linear fit (in logarithmic space) of the RMS of the relative energy error starting
  from $t=1000\,\textnormal{y}$.\label{fig:trappistenergy}}
\end{figure}

As in \S\ref{subsec:nbody}, we operate in double precision with a tolerance of $\epsilon = 10^{-18}$, and we activate high accuracy mode. Since the number of bodies is not prohibitively high, we do not activate compact mode. Some experiments using $\epsilon = 10^{-9}$ are also reported.
We generate a set of 128 initial conditions drawing from the posterior uncertainties reported by \citet{agol2020refining} in terms of the planets osculating Jacobi elements $P_i, T_i, e_i\cos\omega_i, e_i\sin\omega_i$ their masses $m_i$ and the star mass $M_\star$. First we draw samples for the planets masses and the stellar mass. Then we draw a sample of Jacobi elements and we make use of the following equations:
\begin{equation}
\begin{aligned}
P_i &= 2\pi\sqrt{\frac{a_i^3}{G(M_\star+\tilde m_i)}},\\
T_i &= t_0 + \frac{P_i}{2\pi} \left( M_i^T-M_i\right),
\end{aligned}
\label{eq:obs2ic}
\end{equation}
to work out the orbits semi-major axes and mean anomalies. The Jacobi's mass $\tilde m_i$ includes also all the masses of bodies with smaller semi-major axis. Following \citet{grimm2018nature} we set all longitudes of the ascending nodes to zero and inclinations to $\pi / 2$. This allows us to compute the mean anomaly at transit $M_i^T$ from the true anomaly $\nu_i^T = \pi / 2 - \omega_i$.
The eccentricities $e_i$ and the arguments of perihelion $\omega_i$ are derived directly inverting the corresponding Jacobi's elements. Finally, the simulation start epoch $t_0=7257.93115525$ days is that  reported in  \citet{agol2020refining}. 
First, we compute the MEGNO chaos indicator for the set of initial conditions generated and over a period of 100 years, obtaining an average value of $\sim 14$ which confirms chaotic dynamics is to be expected.
We then benchmark our integrator over a simulation of $10^3$ years against  \texttt{IAS15} \citep{rein2015ias15} and  \texttt{WHfast} \citep{rein2015whfast}, using two different time step for \texttt{WHfast}. 
The results are shown in Table \ref{tab:trappist_comparison} where we find, as expected and consistently to Figure \ref{fig:ss_maker_00}, excellent CPU performances at all accuracies. 
In particular, it is clear how the use of our Taylor integrator is very competitive with respect to state of the art approaches for the integration of planetary systems showing superior performances for  the TRAPPIST-1 case here considered. 
We must, though, note that for longer periods of integration the relative error on the Energy will eventually follow Brouwer's law and thus grow, while for symplectic integrators such as \texttt{WHfast} it is likely to stay at the same level. 
We thus  perform a CPU intensive 10My long simulation, corresponding to $\sim  2.4\cdot 10^{9}$ orbits of the inner planet, from all initial conditions sampled and using $\epsilon = 10^{-18}$. 
We find a stable evolution of TRAPPIST-1 in all cases, thus confirming the 
results obtained by \citet{agol2020refining}.
We define the stability of a particular simulation by checking that the semi-major axis of all TRAPPIST-1 planets fluctuate within $10^{-2}$AU.
In Figure \ref{fig:trappistenergy} we show the averaged trend of the relative energy error which follows, as expected, the Brouwer's law and ends up at an average value around $10^{-11}$ allowing to confirm the long term stability of the TRAPPIST-1 system in a high accuracy setting.

\begin{table}
\caption{Accuracy and runtime performances to integrate the TRAPPIST-1 system for $10^{3}$ years from one generated initial condition. }
\label{tab:trappist_comparison}
\centering
\begin{tabular}{lcc}
\toprule
& CPU time (s) & Rel. energy error \\
\midrule
\heyoka{} ($\epsilon=10^{-18}$)& 114 & $3.2\cdot 10^{-15}$   \\
\heyoka{} ($\epsilon=10^{-9}$)& 24 & $1.09\cdot 10^{-8}$   \\
\texttt{WHfast} ($3\cdot 10^{-2}$ days) & 29  &  $1.23\cdot 10^{-8}$  \\
\texttt{WHfast} ($3\cdot 10^{-3}$ days) & 209 & $8.9\cdot 10^{-10}$  \\
 \texttt{IAS15} & 215 & $1.4\cdot 10^{-15}$ \\

\bottomrule
\end{tabular}
\end{table}

\subsection{Batch mode benchmarks}
\label{subsec:batch_benchs}
As we mentioned in \S\ref{subsec:batch}, our integrator is able to operate in batch mode. In batch mode the integrator uses SIMD instructions, available on modern CPUs,
to propagate simultaneously multiple trajectories. All 64-bit x86 CPUs support the SSE2 instruction set, which provides the ability to operate on double-precision
SIMD vectors of size 2. Most CPUs sold from 2011 onwards also support the AVX instruction set, which provides instructions capable of operating on double-precision
SIMD vectors of size 4.

We need to emphasize that the use of batch mode allows to increase the floating-point \emph{throughput} of our integrator, but not its \emph{latency}. In other words,
an integration in batch mode requires (roughly) the same amount of time to complete as in scalar mode, the difference being that in batch mode multiple trajectories
are propagated at the same time. Thus, as we mentioned in \S\ref{subsec:batch}, batch mode is most useful in scenarios where the same dynamical system needs
to be integrated using a variety of different initial conditions or parameters.

In order to evaluate the effectiveness of batch mode, we use the same setup presented in \S\ref{subsec:nbody}, that is, a high-accuracy integration of the
outer Solar System. Each trajectory in the batch is integrated starting from a slightly altered version of the initial conditions reported
in \citet{applegate1986outer}. The total integration time is set to $10^6$ years, and we test batch sizes of 2 (for SSE2) and 4 (for AVX). We measure
the floating-point throughput of the integration by taking the total CPU time and then dividing it by the batch size, and we compare it
to the throughput of an integration in scalar mode.

\begin{figure}
  \centering
   \includegraphics[width=1\columnwidth]{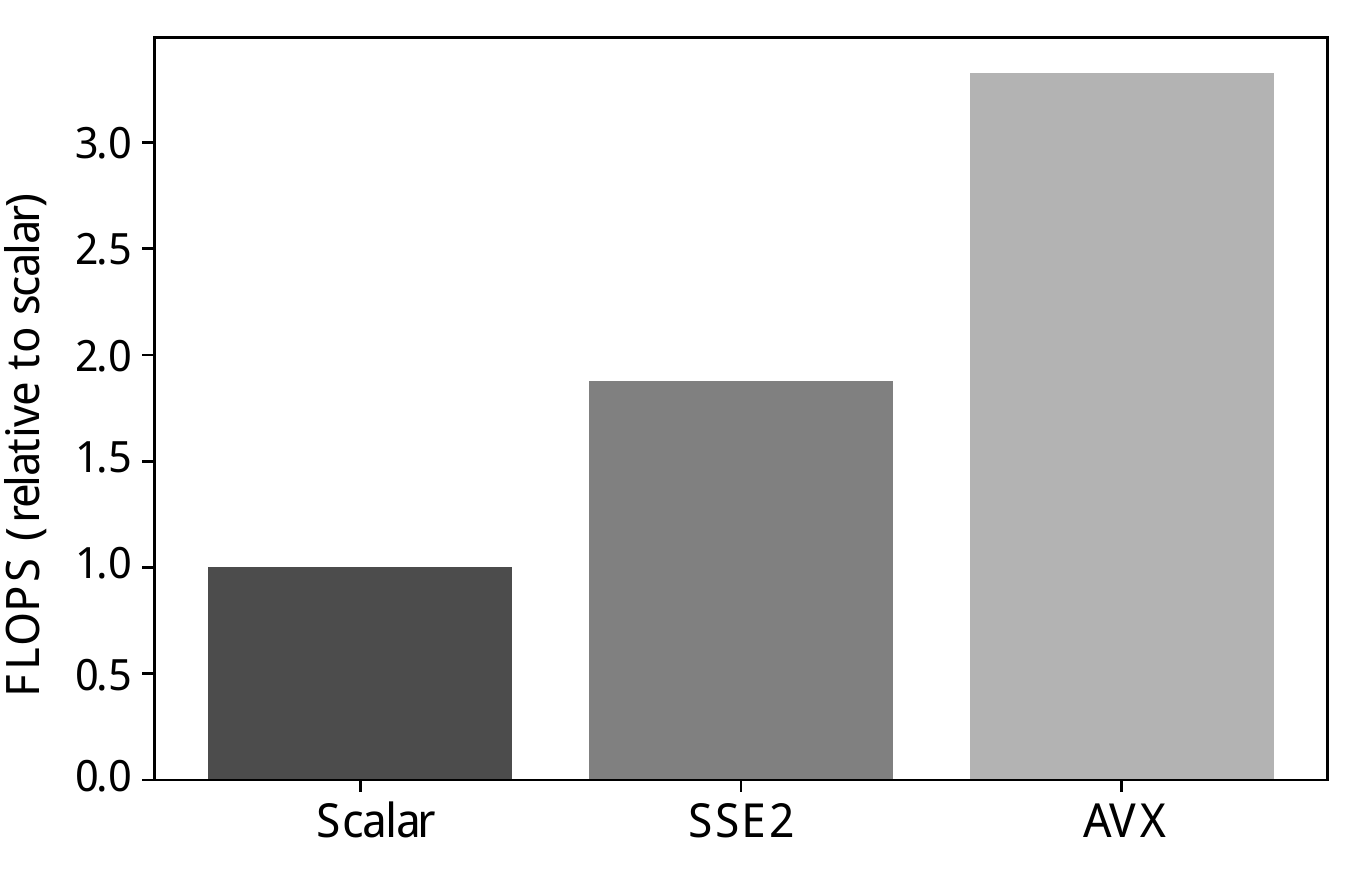}
  \caption{FLOPS comparison between scalar and batch modes using different SIMD instruction sets. The measurements refer to an integration of the outer Solar System
  for $10^6$ years. In batch mode our integrator is propagating multiple trajectories at the same time (2 when using SSE2, 4 when using AVX). With SSE2, the throughput
  increase is linear ($\sim 2\times$), while with AVX the throughput increase is $\sim 3.3\times$ (i.e., $\sim 82\%$ of the theoretical maximum).}
  \label{fig:batch_perf_00}
\end{figure}

The results of the comparison are visualised in Figure \ref{fig:batch_perf_00}. The plot shows how the increase in throughput over a scalar
integration is close to optimal ($\sim 2\times$)
when using SSE2, while when using AVX the throughput increase is $\sim 3.3\times$ (compared to a theoretical maximum of $4\times$).

\subsection{Extended precision integrations}
All the tests and examples shown so far employ double-precision arithmetic. However, as we mentioned in \S\ref{sec:comp_high_acc},
one of the important features of Taylor methods is their efficiency at very low error tolerances, which can be reached via the use
of extended-precision floating-point datatypes (see \S\ref{subsec:supp_fp_types}). Although traditionally double precision
has been considered enough for high-precision applications in astrodynamics and celestial mechanics, in recent times
our understanding of the physics of the Solar System has progressed to the point that the use of double-precision arithmetic can become
the limiting factor in achieving the desired level of accuracy \citep{laskar2011la2010}.

In this last example, we repeat the outer Solar System integration presented in \S\ref{subsec:nbody} in 80-bit IEEE extended precision and IEEE quadruple precision.
In order to account for the use of extended precision datatypes, we set the tolerances respectively to $10^{-21}$ and $10^{-36}$,
and we limit the total integration time to $10^7$ years. As in \S\ref{subsec:nbody}, for each extended precision
datatype we run multiple integration instances in which the initial conditions are slightly and randomly perturbed versions of the values
reported in \citet{applegate1986outer}.

\begin{figure}
  \centering
   \includegraphics[width=1\columnwidth]{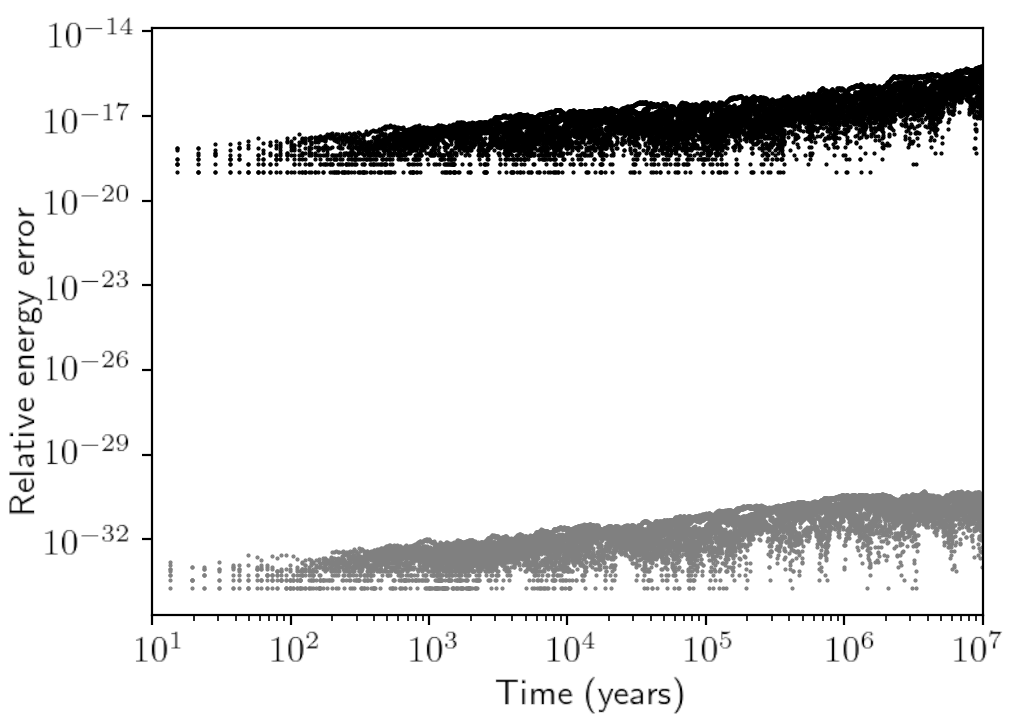}
  \caption{Relative energy error in extended-precision integrations of the outer Solar System. The setup is the same described
  in \S\ref{subsec:nbody}. The black dots refer to integrations in 80-bit IEEE extended precision ($\sim 19$ decimal digits), while
  the gray dots refer to integrations in IEEE quadruple precision ($\sim 34$ decimal digits). Like in \S\ref{subsec:nbody}, we initialise
  several instances of the outer Solar System with initial conditions slightly perturbed with respect to those reported in \citet{applegate1986outer},
  and we integrate them for $10^7$ years.}
  \label{fig:outer_ss_ep_00}
\end{figure}

The results of the integrations are visualised in Figure \ref{fig:outer_ss_ep_00}, where we plot over time the relative energy error
for the 80-bit (black) and quadruple-precision (gray) setups. The plot shows how our Taylor integrator is indeed capable
of achieving Brouwer's law in extended precision.
From the performance point of view, the runtimes for this benchmark, measured on an Intel Xeon Gold 6226, are
$\sim \SI{64}{\second}$ for the 80-bit integration and $\sim \SI{10500}{\second}$ for the quadruple-precision integration
(vs $\sim \SI{13}{\second}$ for a double-precision integration). As mentioned in \S\ref{subsec:supp_fp_types}, quadruple-precision
floating-point types are emulated in software on Intel CPUs, and thus performance is markedly worse with respect to hardware floating-point
types.

\section{Limitations and potential improvements}
\label{sec:limitations}
Despite the fact that our tests and numerical experiments show that Taylor integrators compare favorably to both
symplectic and non-symplectic integrators for high-precision astrodynamical applications, we need to point out
a few drawbacks and limitations that make Taylor integrators not necessarily the best choice in all situations.

\subsection{Differentiability}
The first obvious drawback of Taylor integrators is that they require the ODE system to be expressed in terms
of differentiable expressions amenable to automatic differentiation techniques. The requirement
on differentiability prevents the use of Taylor integrators when the right-hand side of the ODE system contains
black-box functions (which instead pose no issues in integrators which necessitate only the evaluation
of the right-hand side of the ODE system). As a concrete example, Taylor methods are not immediately
usable in conjunction with tree-based gravity computation methods \citep{barnes1986hierarchical},
which do not provide a differentiable formula for the gravitational acceleration.

\subsection{Storage requirements}
The process of automatic differentiation described in \S\ref{subsec:AD} requires an amount of memory storage which
is directly proportional to the complexity of the expressions appearing in the right-hand side of the ODE
system. This means that, for instance, in N-body problems a Taylor integrator requires
$\textnormal{O}\left( N^2 \right)$ memory, whereas in specialised N-body integrators the memory
requirement is only $\textnormal{O}\left( N \right)$.

While, as we showed in our numerical tests,
this is not an issue in practice for $N \lesssim 50$, for higher values of $N$
the memory requirement might impact the usability of Taylor integrators in N-body problems. In addition
to the obvious memory exhaustion problem, another practical issue is the pressure on the memory
subsystem: for high numbers of bodies the amount of memory required by a Taylor integrator will not fit
any more in the processor's cache, which may have a severe impact on the overall runtime performance.
On the other hand, however, with the help of cache-conscious algorithms and data structures,
a carefully-coded implementation specialised for N-body problems may be able to overcome these memory issues.

\subsection{General-purpose vs dedicated}
Our Taylor integrator implementation is generic, in the sense that the right-hand side
of the ODE system can contain arbitrary differentiable expressions. While, on the one hand, this allows
for great flexibility, on the other hand this genericity can also result in suboptimal
performance. Specifically, in ODE systems in which the evaluation of the right-hand side
dominates the runtime, it is often possible for specialised integrators to optimise the
computation in ways that are out of reach for a general-purpose integrator like ours.

In order to give a concrete example, let us consider again the integration
of N-body problems. For low values of $N$, the most efficient way to compute
the bodies' mutual gravity is the usual $\textnormal{O}\left( N^2 / 2 \right)$
loop in which the gravity exerted by particle $i$ on particle $j$ is re-used
to compute the gravity of particle $j$ on particle $i$. Additionally,
in specialised N-body integrators it is also customary to pre-compute
and re-use various factors within the inner part of the loop in order
to minimise the amount of floating-point operations. Whereas our integrator
is able to avoid computing the mutual $ij$ interaction twice, it is currently
unable to pre-compute and re-use common factors in the same way a specialised
integrator can. Thus, in N-body problems, our integrator ends up performing more floating point
operations than optimally necessary.

This sub-optimality ultimately stems from the fact that currently the only way our
implementation can optimise the evaluation (and differentiation) of the right-hand
side of the ODE is via a process of automated common subexpression elimination. A specialised
integrator, on the other hand, can perform more targeted and domain-specific optimisations
when decomposing the ODE system in elementary subexpressions.

\section{Conclusions}
In this paper, we have described and studied in detail \heyoka{}, a new, modern implementation of Taylor's method for the numerical
integration of ODEs. Despite its appealing features, Taylor's method is not widely adopted in the
astrodynamical research community,
in large part due, we believe, to implementation challenges and the suboptimal ergonomics of existing packages. Our implementation
is easy to use from both the C++ and Python languages and, contrary to existing implementations, it does not require the use
of preprocessors or translators, relying instead on a just-in-time compilation approach. Innovative features of our implementation
include support for extended precision and the capability to exploit SIMD instructions in modern processor (which
can lead to a substantial throughput increase).

Our Taylor integrator was tested on a variety of challenging applications in astrodynamics and celestial mechanics. In the context
of long-term N-body integrations, we have shown  how Taylor integrators compare favorably to both symplectic and
non-symplectic integrators, both in terms of runtime performance and capability to maintain high precision. In particular, numerical experiments
set in the outer Solar System and in the exoplanetary system TRAPPIST-1 indicate that our implementation
is able to respect Brouwer's law for the conservation of energy over billions
of dynamical timescales.

In a simulation of the formation of Kirkwood's gaps in the asteroid belt, we reproduced known results regarding the
dynamical lifetimes of asteroid populations in the gaps, thus showing how our implementation is capable of accurately
simulating the close planetary flybys necessary to extract asteroids from the resonances. The ability to handle close encounters
was further confirmed when we simulated Apophis' flyby of the Earth, projected to occur in 2029. Interestingly, in this setup, our general-purpose
integrator outperformed existing domain-specific methods while using a simple Cartesian description of the
system state and a general-purpose adaptive timestepper.

Simulations of gravitational dynamics around highly-irregular extended bodies represented as mascon models
have shown how our implementation of Taylor's method is capable of dealing with ODEs consisting of tens of thousands
of terms. Even in this setup, our integrator showed superior performance with respect to the numerical integration
schemes normally adopted for these studies.

The results of these numerical experiments show how our general-purpose implementation of Taylor's method
compares favourably to a variety of targeted, domain-specific methods in several difficult gravitational problems.
Our integrator does not require tuning or warmup, and it features a single-step general-purpose adaptive timestep
deduction scheme which performs well both in long-term integrations
and in the simulation of close encounters.

We hope that our work will stimulate practitioners in astrodynamics and celestial mechanics to consider the use of Taylor methods,
and also to produce further research on an approach that may enable new significant advances in computational celestial mechanics.
\heyoka{} is available as an open-source project at \url{https://github.com/bluescarni/heyoka}.

\section*{Acknowledgement}
This work is supported by the Deutsche Forschungsgemeinschaft (DFG, German Research Foundation) under Germany's
Excellence Strategy EXC 2181/1 - 390900948 (the Heidelberg STRUCTURES Excellence Cluster).

We would like to thank the anonymous reviewers for helpful comments and suggestions
that improved the quality of the paper.

\section*{Data Availability}
No new data were generated or analysed in support of this research. The source code of the software described in the paper
is freely available under an open-source license at the repository \url{https://github.com/bluescarni/heyoka}.



\bibliographystyle{mnras}
\bibliography{astro_taylor} 


\bsp	
\label{lastpage}
\end{document}